\pdfoutput=1
\documentclass[aps,prb,superscriptaddress,nofootinbib,10pt,a4paper,longbibliography]{revtex4-2}
\usepackage{amsmath,amssymb,amsfonts,graphicx,bm,color}
\usepackage[colorlinks=true,allcolors=blue]{hyperref}
\usepackage[caption=false]{subfig}
\usepackage{tikz}
\usetikzlibrary{decorations.pathreplacing}

\newcommand{\beq}{\begin{eqnarray}}
\newcommand{\eeq}{\end{eqnarray}}

\DeclareMathOperator{\SO}{SO}

\newcommand{\p}{\partial}
\renewcommand{\i}{\mathrm{i}}
\renewcommand{\d}{\mathop{}\!\mathrm{d}}

\DeclareMathOperator{\arctanh}{arctanh}

\newcommand{\calQ}{\mathcal{Q}}

\newcommand{\bx}{\mathbf{x}}

\newcommand{\dE}{\delta{\mkern-1.5mu}E}
\newcommand{\dEwt}{\delta{\mkern-1.5mu}\widetilde{E}}
\newcommand{\dbn}{\delta{\mkern-0.5mu}\mathbf{n}}

\newcommand{\bB}{\mathbf{B}}
\newcommand{\bH}{\mathbf{H}}

\newcommand{\bd}{\mathbf{d}}
\newcommand{\be}{\mathbf{e}}
\newcommand{\bn}{\mathbf{n}}

\newcommand{\figsfolder}{figs/}

\begin{document}
\title{Creation of domain-wall skyrmions in chiral
  magnets with Landau-Lifshitz-Gilbert dynamics and demagnetization}
\date{April, 2024}
\author{Sven Bjarke Gudnason}
\email[Corresponding author: ]{gudnason(at)henu.edu.cn}
\affiliation{Institute of Contemporary Mathematics, School of
  Mathematics and Statistics, Henan University, Kaifeng, Henan 475004,
  P.~R.~China}
\affiliation{Department of Physics, Chemistry and Pharmacy,
  University of Southern Denmark, Campusvej 55, 5230 Odense M,
  Denmark}
\author{Yuki Amari}
\affiliation{International Institute for Sustainability with Knotted
  Chiral Meta Matter (WPI-SKCM$^2$), Hiroshima University, 1-3-1 Kagamiyama,
  Higashi-Hiroshima, Hiroshima 739-8531, Japan}
\affiliation{Research and Education Center for Natural Sciences, Keio
  University, 4-1-1 Hiyoshi, Yokohama, Kanagawa 223-8521, Japan}
\affiliation{Department of Physics, Keio University, 4-1-1 Hiyoshi,
  Yokohama, Kanagawa 223-8521, Japan}
\author{Muneto Nitta}
\affiliation{Department of Physics, Keio University, 4-1-1 Hiyoshi,
  Yokohama, Kanagawa 223-8521, Japan}
\affiliation{Research and Education Center for Natural Sciences, Keio
  University, 4-1-1 Hiyoshi, Yokohama, Kanagawa 223-8521, Japan}
\affiliation{International Institute for Sustainability with Knotted
  Chiral Meta Matter (WPI-SKCM$^2$), Hiroshima University, 1-3-1 Kagamiyama,
  Higashi-Hiroshima, Hiroshima 739-8531, Japan}
\begin{abstract}
Absorption of an isolated bulk magnetic skyrmion into an empty domain
wall in a chiral ferromagnetic system is studied using the
Landau-Lifshitz-Gilbert equation with and without the demagnetization
effect taken into account.
The full phase diagram of creation versus repulsion or annihilation is
mapped out in case of both Bloch-type and N\'eel-type DMI, with and
without demagnetization.
Finally, the unstable domain wall, realizable with a setup of several
external magnets, contains the theoretical possibility of producing a
1-dimensional version of the Kibble-Zurek mechanism, which in turn can
create a number of skyrmion-anti-skyrmion pairs engulfed in the domain
wall: We denote them domain-wall-skyrmion-anti-domain-wall-skyrmion
pairs. 
\end{abstract} 
\maketitle

\section{Introduction}

Magnetic skyrmions are topological excitations of the magnetization
vector in chiral ferromagnetic materials which are quasi-particles in
thin magnetic films and magnetic skyrmion strings in thicker
ferromagnets
\cite{1989JETP...68..101B,1995JETPL..62..247B,Nagaosa2013}.
Their topological stability as well as the fact that they can be moved
with much smaller currents and energy consumption than domain lines,
make them promising candidates for future data storage technological
applications, like race-track memory devices \cite{Fert2013} (for a
review see Ref.~\cite{Back:2019xvi}). 
Magnetic skyrmions have been realized experimentally in MnSi
\cite{doi:10.1126/science.1166767} as well as in
$\text{Fe}_x\text{Co}_{1-x}\text{Si}$ \cite{Yu2010}, and are therefore
not just theoretical topological solitons, but experimentally
realizable pseudo particles with a theoretical nonlinear sigma model
description\footnote{Nonlinear sigma models were first invented to
describe strong interaction in the pion sector, but are mathematically
identical to hard ferromagnets at the level of energy functionals,
without taking into account the dynamics. }.
The nonlinear constraint comes from the fact that hard ferromagnets
are described by the magnetization vector of a constant length called
the magnetic saturation density.

The topology of the magnetic skyrmion is due to the plane (in case of
thin films) or a cross section (in case of thicker materials) being
mapped to a sphere, which is the degrees of freedom of the
3-dimensional magnetization vector with a fixed length.
The topology is, nevertheless, strictly speaking only mathematically
correct for infinitely extended magnetic materials.
Once there is a boundary, the skyrmions may unwrap or nucleate.

Another soliton pertinent to chiral ferromagnets is the domain line,
in case of thin films, or domain walls (DWs), in case of the thicker
materials\footnote{We will from hereon after not distinguish these two
and simply call them DWs. }.
The topology of the DW depends on the specific
potential.
The DW exists and is stable in the presence of the easy-axis
anisotropy potential \cite{doi:10.1126/science.1145799,KUMAR20221},
which is the one we will consider in this paper, in particular we will
consider the out-of-plane anisotropy.

The magnetic skyrmion, although topologically protected in
ferromagnets, is not guaranteed to be stabilized to a finite size:
This is known as Derrick's no-go theorem \cite{Derrick:1964ww}.
For the magnetic skyrmion, luckily, the loophole to Derrick's theorem
is the presence of the Dzyaloshinskii-Moriya interaction (DMI) that is
induced by the spin-orbit coupling (SOC) 
\cite{DZYALOSHINSKY1958241,Moriya:1960zz} (see Appendix
\ref{app:Derrick} for a short review).
It circumvents the shrinking instability, because it only contains a
single derivative and its energy contribution can be negative (which
is necessary for stability of the isolated skyrmion).
The DMI comes in two variants known as the Bloch-type DMI due to the
Dresselhaus SOC and the N\'eel-type DMI due to the Rashba SOC.
The magnetic skyrmion is qualitatively the same for these two types of
DMI, but whereas the magnetization vector curls around the magnetic
skyrmion for the Bloch DMI, it points always in the radial direction
for the N\'eel DMI. 

A great motivation for studying and developing magnetic skyrmions,
magnetic DWs etc., is that racetrack memory devices are envisioned to
be realizable and energy/cost-effective in chiral ferromagnetic
materials.
One of the first prototypes of racetrack memory was the proposal of
magnetic DWs in magnetized nanowires
\cite{doi:10.1126/science.1145799}.
However, the magnetic skyrmions quickly gained large interest in the
community due to the promise of much smaller currents for controlling
the positions of the skyrmions \cite{Fert2013}.

A monkey-wrench on the path to great success for the magnetic skyrmion
to become the ultimate low-energy storage device candidate, is the
skyrmion Hall effect \cite{Zang2011,wanjun2017,chen2017skyrmion}.
Once a current is applied to a magnetic skyrmion it veers off its
forward path: viz.~the skyrmion trajectories bend.
This makes it difficult to control the exact positions of the
skyrmions and if they are to be used as bits in a memory device and
they suddenly moved half a spot to either side, it may be difficult or
impossible to interpret the data as being a one or zero.

As was perhaps clear from the above discussion, the DWs are harder to
move than the skyrmions, and hence a nice idea to pursue is the
possibility to encapsulate the magnetic skyrmion into the DW: This
composite soliton in the plane is called the DW-skyrmion. 
The skyrmion may still move, but its motion is restricted to be along
the line of the DW (in thin film, i.e.~2-dimensions).
In the planar samples of chiral ferromagnetic materials, stable
DW-skyrmions have been proposed
\cite{Kim:2017lsi,PhysRevB.99.184412,PhysRevB.102.094402,Kuchkin:2020bkg,Ross:2022vsa,Lee:2022rxi,Amari:2023gqv,Amari:2023bmx}.\footnote{DW-skyrmions
have been previously studied also in field theory in two dimensions
\cite{Nitta:2012xq,Kobayashi:2013ju,Jennings:2013aea,Bychkov:2016cwc}
and in three dimensions
\cite{Nitta:2012wi,Nitta:2012rq,Gudnason:2014nba,Gudnason:2014hsa,Eto:2015uqa,Nitta:2022ahj},
and have recently been found to support a new phase in QCD 
\cite{Eto:2023lyo,Eto:2023wul,Eto:2023tuu,Amari:2024mip,Amari:2025twm}.
}
The DW-skyrmion, like the magnetic skyrmion, has already been
discovered in the laboratory
\cite{Nagase:2020imn,10.1063/5.0056100,Yang2021}.
Moreover, dynamics of DW-skyrmions \cite{PhysRevB.109.014404,Chen_2025,Nie_2025,PhysRevB.111.024402,b2s4-d69w,chen2025currentinduceddynamicsblochdomainwall} and magnonic excitations on DW-skyrmions \cite{Saji_2025,wang2025domainwallskyrmionbasedmagnonic} have been studied.
The DW-skyrmions in chiral soliton lattices, however, are unstable and 
decay into two merons \cite{Amari:2023bmx}, which are quasi-particles
with topological charge $1/2$
\cite{Muller2017,Kharkov2017,Gobel2019,Mukai2022}. 
Skyrmions located in a domain-wall junction were also studied \cite{Lee:2024lge}.

In this paper, we continue the study of the creation of DW-skyrmion(s)
from a bulk magnetic skyrmion and an empty DW, which we started in
Ref.~\cite{Gudnason:2024shv}.
Here, we utilize the more physical Landau-Lifshitz-Gilbert equation
for the dynamical evolution of initial states, as opposed to the
energy minimization techniques used in Ref.~\cite{Gudnason:2024shv}
(energy minimizing techniques like gradient flow correspond
to large (infinite) Gilbert damping parameter).
Furthermore, we consider here both the Bloch-type and the N\'eel-type
DMI with and without the magnetization effect taken into account.
Although for the Bloch DMI the demagnetization field is trivial for
both the DW and the magnetic skyrmion, it is nontrivial for the
DW-skyrmion.
For the N\'eel DMI, instead, the demagnetization field is always
nontrivial and affects the solitons at hand.
For the DW and the magnetic skyrmion, the effect is simply a reduction
of size, whereas for the DW-skyrmion it is more complicated. 
We find the full phase diagrams for creation of DW-skyrmions for all
four cases of Bloch and N\'eel DMI with and without demagnetization
(without demagnetization the two DMIs are equivalent though).
Furthermore, we explore in detail the most unstable DW which gives rise to
a 1-dimensional Kibble-Zurek mechanism 
\cite{Kibble:1976sj,Kibble:1980mv,Kibble:1980mv,Zurek:1985qw} 
which can create any number of
DW-skyrmion-anti-DW-skyrmions.
We dub the angle or line in the phase diagram where this occurs
\emph{the Kibble line}.
We also calculate effective Thiele equations for the movement of the
unstable DW in both cases of Bloch- or N\'eel-type DMI.
The results of the Thiele equation or moduli space approximation, are
in very good accord with the full numerical computations using the LLG
equation. 
Finally, we provide simulation videos in the ancillary files of
a large number of initial states, which are all marked with labels in
the phase diagrams.

The paper is organized as follows.
In Sec.~\ref{sec:model}, we introduce the model and set the notation
for the paper.
In Sec.~\ref{sec:Bloch_vs_Neel}, we demonstrate, in our notation, 
the well-known fact that Bloch DMI and N\'eel DMI are equivalent
without demagnetization but are inequivalent with the demagnetization
effect taken into account.
In Sec.~\ref{sec:constituent_solitons}, we review the magnetic
skyrmion and DW in the stereographic Riemann coordinate, show that the
demagnetization field is trivial in the Bloch DMI case and take the
demagnetization effect into account in the N\'eel DMI case, which
essentially gives rise to smaller solitons or alternatively a larger anisotropy coefficient.
In Sec.~\ref{sec:initial_cond}, we set up the initial states for the
numerical computations leading to the phase diagrams of whether the
bulk magnetic skyrmion can be absorbed into the empty DW or not.
The numerical method is then explained in Sec.~\ref{sec:num_method}.
Sec.~\ref{sec:results} presents the numerical results, including the
phase diagrams and examples of the Kibble-Zurek mechanism at work.
Finally, we conclude with a discussion and some outlook on future work
in Sec.~\ref{sec:conclusion}. 
We delegate the brief discussion of using random noise as a trigger
for the Kibble-Zurek mechanism to Appendix \ref{app:random},
a review of Derrick's theorem to Appendix \ref{app:Derrick} and
finally the monitoring of the topological charge as a function of
time for selected LLG flows to Appendix \ref{app:topo_charge}.

\section{The chiral magnetic model with Landau-Lifshitz-Gilbert dynamics}\label{sec:model}

In this paper, we will consider the energy functional comprised by the
Heisenberg exchange energy, the Dzyaloshinskii-Moriya-Interaction
(DMI), an easy-axis anisotropy potential as well as the
demagnetization energy
\beq
E = T\int\left[A\,\p_i\bn\cdot\p_i\bn
  + D\bn\cdot\bd_i\times\p_i\bn
  + K (1-n_3^2)
  - \frac{\mu_0 M_{\rm sat}}{2}\bn\cdot\bH_{\rm demag}
  \right]\d^2x,
\label{eq:E}
\eeq
where $T$ is the material thickness,
$A$ is the exchange stiffness constant,
$\bn=(n_1,n_2,n_3)$ is the unit magnetization vector ($\bn\cdot\bn=1$)
defined by $\bn=\mathbf{m}/M_{\rm sat}$,
$\mathbf{m}$ is the unnormalized magnetization vector,
$M_{\rm sat}$ is the magnetic saturation density,
$\p_i$ is the partial derivative in the plane,
$i=1,2$ with repeated indices summed over by the Einstein convention, 
$D$ is the DMI coupling,
$\bd_i$ are two vectors parametrizing the type of the DMI,
$K$ is the anisotropy constant,
and $\bH_{\rm demag}$ is the demagnetization field.
We assume that the chiral magnet is a thin film, so there is
essentially no dependence on the third spatial coordinate: $\p_3=0$
and $\bn=\bn(x_1,x_2)$.

We will consider only the case without currents, so the Maxwell
equations are simply $\nabla\cdot\bB=0$ and $\nabla\times\bB=0$ with
$\nabla=(\p_1,\p_2,0)$ (in the plane).
Usually, the situation with a current forces one to choose a vector
potential for the magnetic field, so that the first equation is solved
manifestly (divergence free field) and the curl can be nonvanishing.
For ferromagnetic systems, however, it is convenient to use a scalar
magnetic potential, which is manifestly curl free and the divergence
free criterion is solved by setting \cite[Chap.~5.9]{JacksonEM} (see
also Refs.~\cite{Qin_2018,Leask:2025pdz}):
\beq
\nabla\cdot\bH_{\rm demag}
= -\nabla\cdot\mathbf{m},
\eeq
with $\bB=\mu_0(\bH+\mathbf{m})$ and $\mu_0$ is the magnetic
permeability of the vacuum.
We choose the scalar magnetic field as
$\bH_{\rm demag}=-M_{\rm sat}\lambda\nabla\Phi$ for which the
above equation becomes simply the magnetostatic Poisson equation
\beq
\nabla^2\Phi = \frac{1}{\lambda}\nabla\cdot\bn,
\eeq
with the divergence of the magnetization field being the
``magnetic charge'' and $\lambda$ is a length unit which we will fix
shortly.

The chiral magnetic system with a Dresselhaus SOC
gives rise to a Bloch-type DMI, whereas the Rashba SOC gives
rise to the Néel-type DMI.
These correspond to fixed vectors as
\beq
\begin{cases}
  \textrm{Dresselhaus SOC (Bloch type)} : & \bd_1=\be_1,\phantom{-}\quad \bd_2=\be_2,\\
  \textrm{Rashba SOC (Néel type)} : & \bd_1=-\be_2,\quad \bd_2=\be_1.
\end{cases}
\label{eq:di_Bloch_Neel}
\eeq

The Landau-Lifshitz-Gilbert (LLG) equation for vanishing currents
reads 
\beq
\p_t\bn = -\gamma\bn\times\bH_{\rm eff} + \alpha_G\bn\times\p_t\bn,
\eeq
where $\gamma$ is called the gyromagnetic ratio, $\alpha_G$ is the
Gilbert damping coefficient, and the effective magnetization is
related to the variation of the free energy by
$\bH_{\rm eff}=-M_{\rm sat}^{-1}\frac{\dE}{\dbn}$.
We can write the LLG equation in the form of the Landau-Lifshitz (LL)
equation, by inserting the LLG equation into itself:
\beq
\p_t\bn = \frac{\gamma}{M_{\rm sat}(1+\alpha_G^2)}\bn\times\frac{\dE}{\dbn}
-\frac{\alpha\gamma}{M_{\rm sat}(1+\alpha_G^2)}\frac{\dE}{\dbn}.
\eeq
where we have used that $\bn\cdot\frac{\dE}{\dbn}=0$ due to the
variational principle, 
$\bn\cdot\p_t\bn=0$ due to the unit-length constraint:
$\bn\cdot\bn=1$ and finally we have rearranged the terms so as to
isolate $\p_t\bn$ obtaining the LLG equation in LL form.

At this stage, the equations are cluttered with physical constants so
it will prove convenient to switch to dimensionless variables and
combine the physical constants into two model parameters.
We denote by $\tilde{x}=x/\lambda$ the dimensionless spatial variable
and $\lambda$ is a length unit; similarly $\tilde{t}=t/\tau$ with
$\tau$ being a time unit.
We thus obtain
\beq
  E = 2TA\widetilde{E},\qquad
  \widetilde{E} = \int\left[\frac12\tilde\p_i\bn\cdot\tilde\p_i\bn
  +\kappa\,\bn\cdot\bd_i\times\tilde\p_i\bn
  +\frac12(1-n_3^2)
  +\eta\,\bn\cdot\widetilde{\nabla}\Phi
  \right]\d^2\tilde{x},\qquad
  \widetilde{E}=\int\mathcal{E}\d^2\tilde{x},
  \label{eq:Etilde}
\eeq
with the LLG and Poisson equations
\begin{align}
\p_{\tilde t}\bn &= \bn\times\frac{\dEwt}{\dbn}
-\alpha_G\frac{\dEwt}{\dbn},\label{eq:LLG}\\
\widetilde{\nabla}^2\Phi &= \widetilde{\nabla}\cdot\bn,\label{eq:Poisson}
\end{align}
where we have fixed the length and time units as
\beq
\lambda = \sqrt{\frac{A}{K}}, \qquad
\tau = \frac{M_{\rm sat}(1+\alpha_G^2)}{2TA\gamma},
\eeq
as well as the model parameters
\beq
\kappa = \frac{D}{2\sqrt{AK}}, \qquad
\eta = \frac{\mu_0 M_{\rm sat}^2}{4K}.
\eeq
The model depends only on three parameters, the effective DMI coupling
$\kappa$, the effective coupling to the demagnetization field $\eta$
and the Gilbert damping coefficient $\alpha_G$ -- all the remaining
constants are just energy, length and time units.

We will choose
\beq
\alpha_G=0.3, \qquad
\kappa=0.4, \qquad
\eta=0.3,
\label{eq:modelparms}
\eeq
which can
be obtained from the following physical constants:
$M_{\rm sat}\sim6\times10^5{\rm A}/{\rm m}$,
$A\sim 10^{-11}{\rm J}/{\rm m}$,
$D\sim1.55\times10^{-3}{\rm J}/{\rm m}^2$,
$K\sim3.75\times10^5{\rm J}/{\rm m}^3$, and
$\mu_0=4\pi\times 10^{-7}{\rm J}/({\rm mA}^2)$.
These values are very close to the parameters
used for micromagnetic simulations of Pt/Co/Ta in
Ref.~\cite{Woo2016} (this reference uses
$D\sim1.5\times10^{-3}{\rm J}/{\rm m}^2$, which is slightly lower than
our value, giving rise to a $\kappa\sim 0.39$ instead of
$\kappa\sim0.4$).

In order to compute the LLG flow, we need the first variation of the
dimensionless energy \eqref{eq:Etilde} which reads
\beq
\frac{\dEwt}{\dbn} = 
-\widetilde\nabla^2\bn
+2\kappa\,\bd_i\times\tilde\p_i\bn
-n_3\,\be_3
+\eta\,\widetilde\nabla\Phi
=0.
\label{eq:eom}
\eeq
In the limit of large $\alpha_G$, the LLG equation becomes just
gradient flow and the negative sign in front of the (2D) Laplacian
operator in the above equation ensures that the energy will relax to a
local minimum.
The phase diagram without demagnetization taken into account
was studied in Ref.~\cite{Gudnason:2024shv}.
The symplectic part of the LLG flow, however, induces transverse
motion affecting the dynamics and altering it from pure gradient
flow.

The magnetic skyrmion carries a topological charge given by
\beq
Q = \int\calQ\;\d^2\tilde{x},\qquad
\calQ = -\frac{1}{4\pi} \bn\cdot\tilde{\p}_1\bn\times\tilde{\p}_2\bn,
\label{eq:Q}
\eeq
which arises by considering the stereographic projection of the plane
to the 2-sphere, so that the magnetization vector is a map from the
2-sphere (the plane) to a unit vector in $\mathbb{R}^3$,
{\it i.e.}~another 2-sphere: this is characterized by
$\pi_2(S^2)=\mathbb{Z}\ni Q$, with $Q$ being the topological charge
above.
Since the quantity is topological, we can use lengths with or without
units obtaining the same result.

\section{Bloch versus Néel}\label{sec:Bloch_vs_Neel}

In this section, we will demonstrate the equivalences and the
differences between the Bloch DMI and the Néel DMI.
We will consider the rotation matrix $R\in\SO(2)\subset\SO(3)$,
i.e.~valued in the $\SO(2)$ subgroup of $\SO(3)$ as
\beq
\bn \to R\bn =
\begin{pmatrix}
  \cos\vartheta&\sin\vartheta&0\\
  -\sin\vartheta&\cos\vartheta&0\\
  0&0&1
\end{pmatrix}\bn,
\eeq
which obviously satisfies $R^{\rm T}R=\mathbf{1}_3$.
The Heisenberg exchange energy and the easy-axis anisotropy potential
are invariant under such a rotation, so we need only examine the DMI
and the demagnetization energy.
Starting with the former, we have
\beq
\kappa (R\bn)\cdot\bd_i\times\tilde\p_i(R\bn)
= \kappa\bn\cdot(R^{\rm T}\bd_i)\times\tilde\p_i\bn,
\label{eq:DMItrans}
\eeq
where we have used that $\det R=1$.
We thus need to find an $R$ that takes the vectors $\bd_i$ of the
Néel-type DMI to the Bloch type, which can easily be seen to be
\beq
R =
\begin{pmatrix}
  0&1&0\\
  -1&0&0\\
  0&0&1
\end{pmatrix},
\label{eq:RBlochNeel}
\eeq
so that
\beq
\bd_i^{\textrm{Bloch}} = R^{\rm T}\bd_i^{\textrm{Néel}}.
\eeq
Now, if we read Eq.~\eqref{eq:DMItrans} from right to left, we can see
that a N\'eel skyrmion $\bn^{\textrm{Néel}}(x,y)$ is formally solved by a
solution $R\bn^{\rm Bloch}(x,y)$ with $\bn^{\rm Bloch}(x,y)$ being a
solution to the chiral magnetic system with a Bloch-type DMI.
In that sense, the solutions look different, but are merely a rotation
of the magnetization vectors by 90 degrees in the $(x,y)$-plane.

Turning now to the demagnetization energy, we can see that already the
Poisson equation \eqref{eq:Poisson} spells trouble since the rotation
(or relabeling) of $\bn$ does not leave $\nabla\cdot\bn$ invariant.
This will change the magnetic scalar potential and this will have
physical consequences, as we shall see shortly.

We have now established that the energy functional \eqref{eq:E} is
invariant under the simultaneous transformation $\bn\to R\bn$ and
$\bd_i\to R\bd_i$ with $R$ given in Eq.~\eqref{eq:RBlochNeel}, where
we can think of $\bd_i$ as the Bloch-type DMI vectors, $\bn$ the
solution to the Bloch equations of motion (stationary point), $R\bd_i$
as the N\'eel-type DMI vectors and finally $R\bn$ as the solution to
the N\'eel system's equations of motion.
One may wonder whether the Bloch and N\'eel systems are equivalent
under the LLG flow.
To answer this, we note that under the transformation $\bn\to R\bn$,
the first variation of the energy functional transforms as
\beq
\frac{\dEwt}{\dbn} \to R\frac{\dEwt}{\dbn},
\eeq
if and only if $\eta:=0$.
In order to show this, one can simply use the invariance of the DMI
term under the transformation $\bn\to R\bn$ and $\bd_i\to R\bd_i$ to
see that
\beq
\bd_i\times\tilde\p_i\bn
\to R(\bd_i\times\tilde\p_i\bn),
\eeq
as the invariance implies that
$(R\bn)\cdot R(\bd_i\times\tilde\p_i\bn)=\bn\cdot\bd_i\times\tilde\p_i\bn$,
just like $(R\bn)\cdot(R\bn)=\bn\cdot\bn$.
This exercise was useful, because we have now shown that the first
term in the LLG equation \eqref{eq:LLG} also transforms as $\bn$
itself: 
\beq
\bn\times\frac{\dEwt}{\dbn} - \alpha_G\frac{\dEwt}{\dbn}
\to R\left(\bn\times\frac{\dEwt}{\dbn} - \alpha_G\frac{\dEwt}{\dbn}\right),
\eeq
exactly like the left-hand side of the LLG equation \eqref{eq:LLG}.

What we have shown in this section, is that for $\eta:=0$ i.e.~without
the demagnetization field coupled to the chiral magnetic skyrmion
system, there is no physical difference between the Bloch and the
N\'eel DMI terms -- both in terms of energies and in terms of LLG
flows.

\section{Constituent solitons}\label{sec:constituent_solitons}

In this section, we will briefly review the constituents and check if
and how the demagnetization field modifies the isolated soliton.
The magnetic skyrmion and domain wall are given by
\begin{align}
  u^{\rm sk} = e^{\i(\theta+\beta)}\tan\left(\frac{f(r)}{2}\right),\qquad
  u^{\rm DW} = e^{\i\alpha}\tan\left(\frac{f(\tilde{x})}{2}\right),\label{eq:u_skyrmion_DW}
\end{align}
the stereographic coordinate is given in terms of the
magnetization vector as
\beq
u=\frac{n_1+\i n_2}{1+n_3},
\eeq
and the polar coordinates are defined as
$\tilde{x}+\i\tilde{y}=r e^{\i\theta}$.

\subsection{Bloch DMI}

We will start with the case of the Bloch-type DMI that comes from the
Dresselhaus SOC, see Eqs.~\eqref{eq:E} and \eqref{eq:di_Bloch_Neel}.

The profile function $f(r)$ of the magnetic skyrmion is not known
analytically, but it can be determined numerically from the equation
\beq
f'' + \frac{f'}{r} - \frac{\sin2f}{2r^2}
+\frac{2\kappa\sin^2f}{r} - \frac{\sin2f}{2}
-\eta\cos(f)\frac{\p_\theta\Phi}{r}
= 0,
\label{eq:eomf_Bloch}
\eeq
which together with
\beq
\cos\beta=0,
\eeq
arise from the first variation of the energy functional \eqref{eq:E},
and the former is the same as the first variation of the energy
functional (in dimensionless units)
\beq
\widetilde{E}^{\rm sk} = \int\bigg[
  (f')^2
  +\sin^2\left(f + \frac{1}{r^2}\right)
  +\kappa\sin\beta\left(2f' + \frac{\sin(2f)}{r}\right)
  +2\eta\sin f\left(\cos\beta\p_r\Phi + \sin\beta\frac{\p_\theta\Phi}{r}\right)
  \bigg]\pi r\d r,
\eeq
due to the principle of symmetric criticality \cite{Palais:1979rca},
see also \cite[p.~258]{Coleman:1985}.
$f'=f'(r)$ and $f''=f''(r)$ denote first- and second-order radial
derivatives, respectively.
$\beta=\pi/2$ minimizes the DMI energy (i.e.~making it maximally
negative, since $f'(r)<0$ for the boundary conditions $f(0)=\pi$ and
$f(\infty)=0$) which in turn switches off the coupling between $f$ and
the radial derivative of the magnetic potential $\Phi$.
Now the equation of motion for $f$ still has a coupling to the angular
derivative of the magnetic potential, which one could think would
alter the magnetic skyrmion.
However, inspecting the Poisson equation \eqref{eq:Poisson}:
\beq
\p_r^2\Phi + \frac{1}{r}\p_r\Phi + \frac{1}{r^2}\p_\theta^2\Phi
= \cos\beta\left(\frac{\sin f}{r} + \cos(f)f'\right),
\label{eq:Poisson_BlochSk}
\eeq
we see that it is unsourced and hence harmonic for $\beta=\pi/2$.
Energy minimization of the harmonic field forces us to set $\Phi:=0$
and so the Bloch skyrmion is unaware of the demagnetization field and
the last term in Eq.~\eqref{eq:eomf_Bloch} vanishes.

We now turn to the Bloch line or Bloch-type DW, whose equation of
motion is given by 
\beq
f'' - \frac12\sin(2f) - \eta\cos(f)\p_{\tilde{y}}\Phi = 0,
\label{eq:eomfDW_Bloch}
\eeq
as well as $\cos\alpha=0$ which again arise from the first variation
of Eq.~\eqref{eq:E} and the above equation for $\eta=0$ is just the
sine-Gordon equation, which also derives from the energy functional
(in dimensionless units)
\begin{align}
  \widetilde{E}^{\rm DW} &=
  \int\left[\frac12(f')^2
    +\frac12\sin^2f
    +\kappa\sin(\alpha)f'
    +\eta\sin f\left(\cos(\alpha)\p_{\tilde{x}}\Phi + \sin(\alpha)\p_{\tilde{y}}\Phi\right)
    \right]\d^2\tilde{x},
  \label{eq:EDW_Bloch}
\end{align}
due to the principle of symmetric criticality and now $f'=f'(\tilde{x})$.
$\alpha=\pi/2$ minimizes the DMI energy (i.e.~making it maximally
negative, since $f'(\tilde{x})<0$ for the boundary conditions $f(-\infty)=\pi$
and $f(\infty)=0$) which leaves only one coupling between $f$ and
the demagnetization field: $\p_{\tilde{y}}\Phi$.
Naively, one would again think that the sine-Gordon equation is
modified by the demagnetization field.
However, inspecting the Poisson equation \eqref{eq:Poisson}:
\beq
\p_{\tilde{x}}^2\Phi + \p_{\tilde{y}}^2\Phi = \cos\alpha\cos(f)f',
\label{eq:Poisson_BlochDW}
\eeq
we see that $\cos\alpha=0$ again makes the Poisson equation unsourced
and hence harmonic.
Energy minimization again leaves us with $\Phi:=0$ everywhere.
Since $\Phi=0$, $f$ is given by the renowned sine-Gordon solution
$f=2\arctan\left(e^{-\tilde{x}}\right)$, corresponding to
\beq
u^{\rm DW} = e^{\i\alpha - (\tilde{x}-X_0)}.
\label{eq:u_DW_sol}
\eeq
The energy (in dimensionless units) then reduces to
\beq
\widetilde{E}^{\rm DW} = Y\left[2 - \pi\kappa\sin\alpha\right],
\eeq
which indeed is minimized by $\alpha=\pi/2$ and $Y$ is the length of the material in the $\be_2$-direction in dimensionless units.

\subsection{Néel DMI}\label{sec:const_soliton_Neel}

We now turn to the magnetic skyrmion and the magnetic DW in the case
of the N\'eel-type DMI that comes from the Rashba SOC, see
Eqs.~\eqref{eq:E} and \eqref{eq:di_Bloch_Neel}.

The profile function $f(r)$ is now determined numerically from
\beq
f'' + \frac{f'}{r} - \frac{\sin2f}{2r^2}
+\frac{2\kappa\sin^2f}{r} - \frac{\sin2f}{2}
-\eta\cos(f)\p_r\Phi
= 0,
\label{eq:eomf_Neel}
\eeq
which together with
\beq
\sin\beta=0,
\eeq
arise from the first variation of the energy functional \eqref{eq:E},
and the reduced energy functional (in dimensionless units) reads
\beq
\widetilde{E}^{\rm sk} = \int\bigg[
  (f')^2
  +\sin^2\left(f + \frac{1}{r^2}\right)
  +\kappa\cos\beta\left(2f' + \frac{\sin(2f)}{r}\right)
  +2\eta\sin f\left(\cos\beta\p_r\Phi + \sin\beta\frac{\p_\theta\Phi}{r}\right)
  \bigg]\pi r\d r.
\eeq
$\beta=0$ minimizes the DMI energy, which leaves the coupling between
$f$ and the radial derivative of the magnetic potential $\Phi$ turned
on. 
The Poisson equation remains that of Eq.~\eqref{eq:Poisson_BlochSk}
and hence the right-hand side is a function of $f$ that sources the
magnetic potential for $\beta=0$.
Assuming that $\Phi=\Phi(r)$ is a radial function only (since the
right-hand side of Eq.~\eqref{eq:Poisson_BlochSk} is a radial
function), we can integrate the Poisson once to get
\beq
\p_r\Phi = \sin f,
\label{eq:Poisson_NeelSk_reduced}
\eeq
which in turn reduces the equation of motion for the profile function
for the N\'eel skyrmion to
\beq
f'' + \frac{f'}{r} - \frac{\sin2f}{2r^2}
+\frac{2\kappa\sin^2f}{r} - \frac{(1+\eta)\sin2f}{2}
= 0.
\label{eq:eomf_Neel_demag}
\eeq
The backreaction of the demagnetization field has the effect of
increasing the ``mass term'' (anisotropy coefficient) in the profile function.
In dimensionless units, the mass is unity and is increased by taking
into account the backreaction of the demagnetization field by a factor
of $\sqrt{1+\eta}$.
Increasing the mass term gives rise to smaller skyrmions.
Since we do not know an analytic solution for $f$, we also cannot
integrate Eq.~\eqref{eq:Poisson_NeelSk_reduced} to obtain an explicit
form for the magnetic scalar potential.
Numerically, however, this first-order equation can easily be
integrated.

Using a scaling argument, the solutions with the increased mass by a factor of
$\sqrt{1+\eta}$ are equivalent to the solutions with unit mass and
$\kappa\to\frac{\kappa}{\sqrt{1+\eta}}$ and
$r\to\frac{r}{\sqrt{1+\eta}}$.
That is, an $\eta=0.3$ corresponds to a decrease in $\kappa$ by
$\sim12\%$ and a decrease in length scales by the same amount
(shrinking of the soliton).

\begin{figure}[!htp]
  \centering
  \mbox{\subfloat[]{\includegraphics[width=0.49\linewidth]{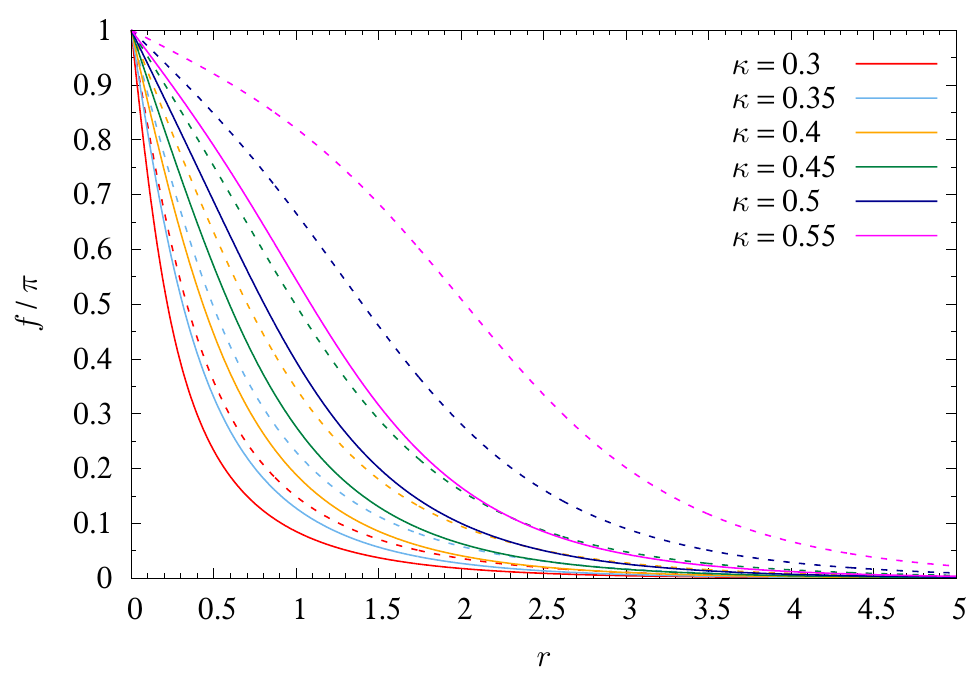}}
    \subfloat[]{\includegraphics[width=0.49\linewidth]{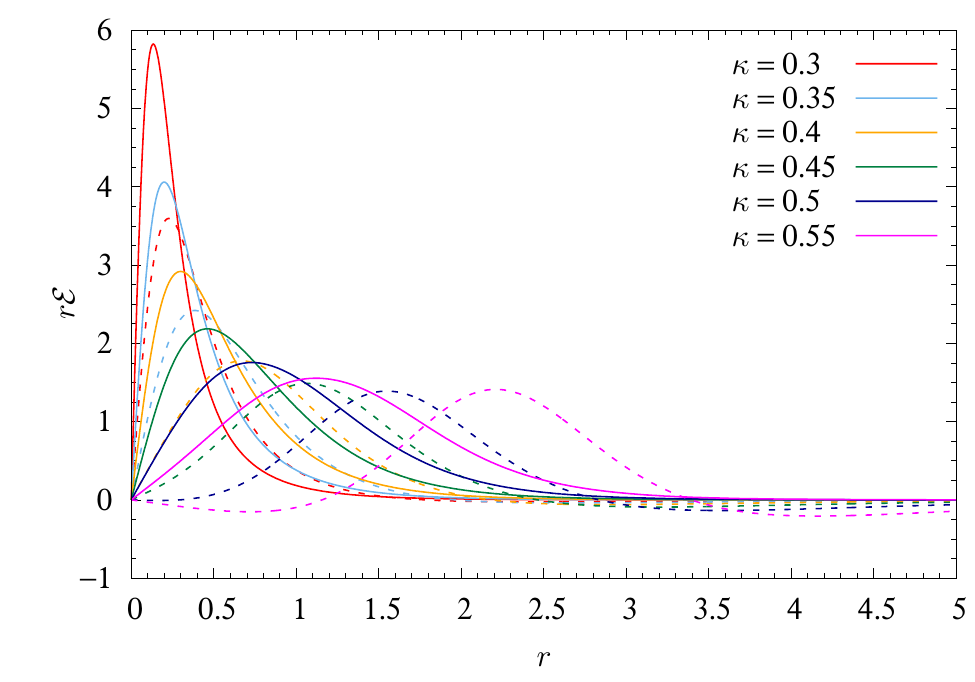}}}
  \mbox{\subfloat[]{\includegraphics[width=0.49\linewidth]{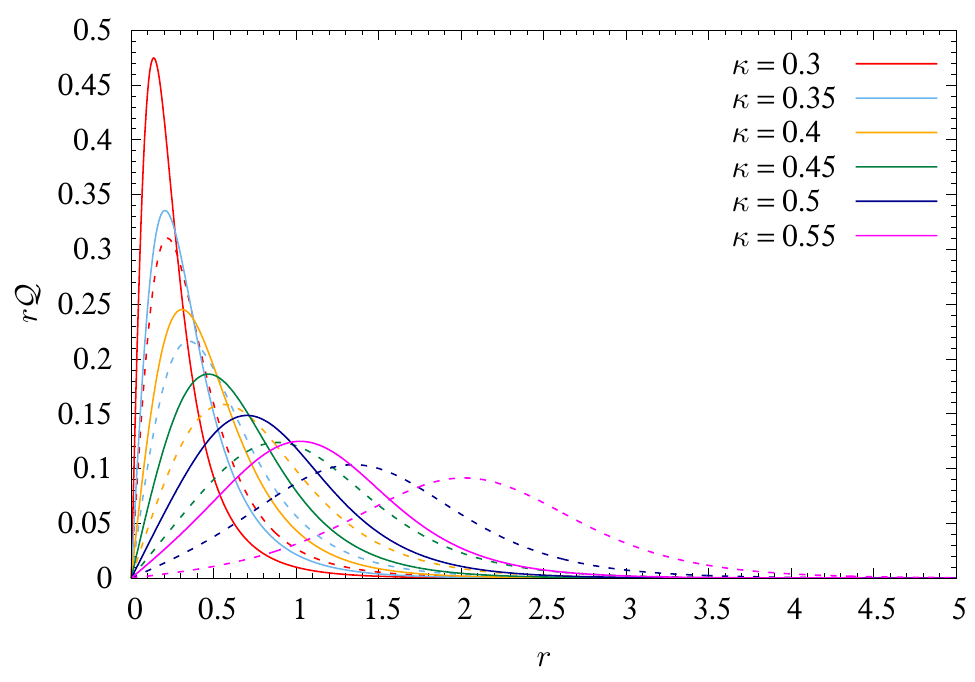}}
    \subfloat[]{\includegraphics[width=0.49\linewidth]{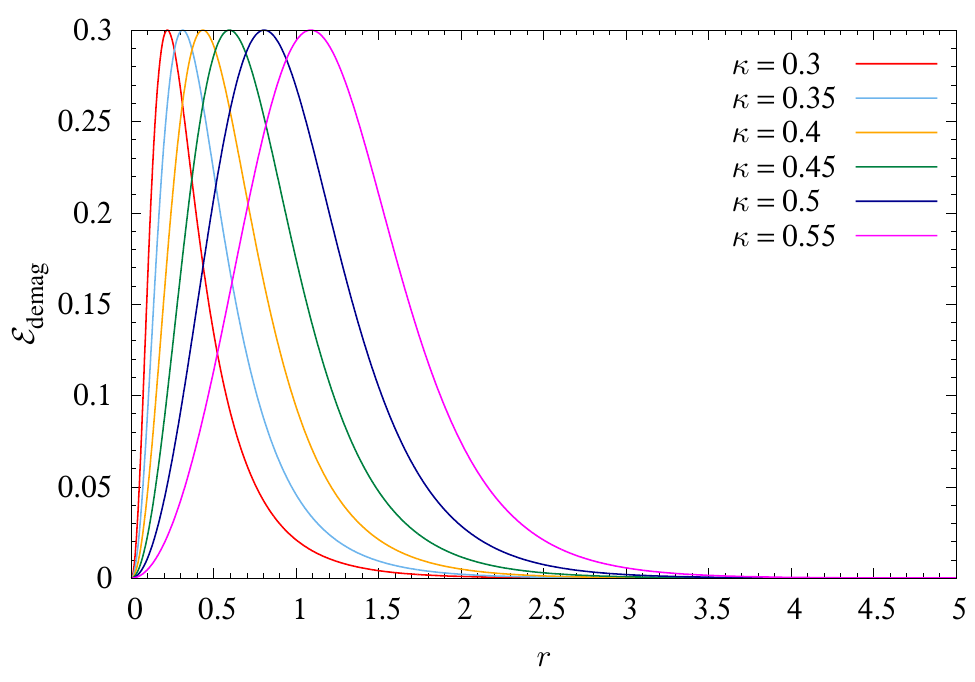}}}
  \caption{(a) profile, (b) energy density, (c) topological charge
    density and (d) demagnetization energy density of the magnetic (N\'eel)
    skyrmion.
    The dashed lines correspond to both Bloch-type and N\'eel-type
    magnetic skyrmions without the demagnetization taken into account
    and the solid lines correspond to the N\'eel-type magnetic
    skyrmions with $\eta=0.3$.
    The DMI coupling is varied from 0.3 to 0.55, and everything is
    plotted in dimensionless units, see the text.
  }
  \label{fig:Neel_demag}
\end{figure}
\begin{figure}[!htp]
  \centering
\mbox{\subfloat[]{\includegraphics[width=0.24\linewidth]{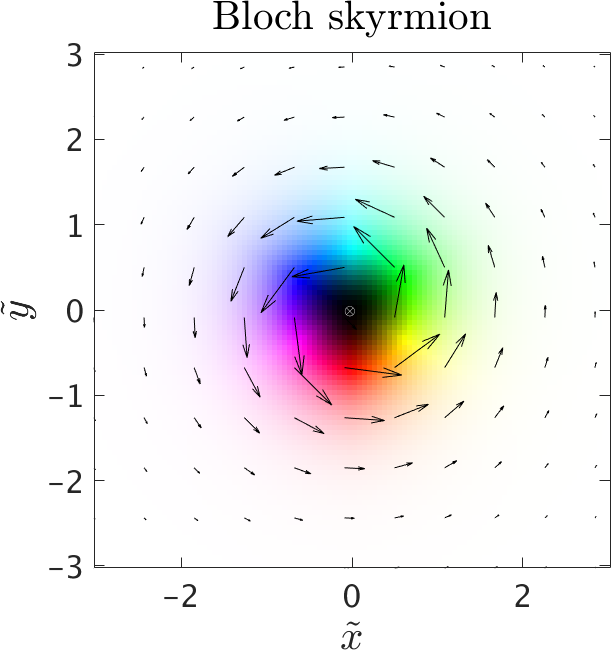}}\ 
    \subfloat[]{\includegraphics[width=0.24\linewidth]{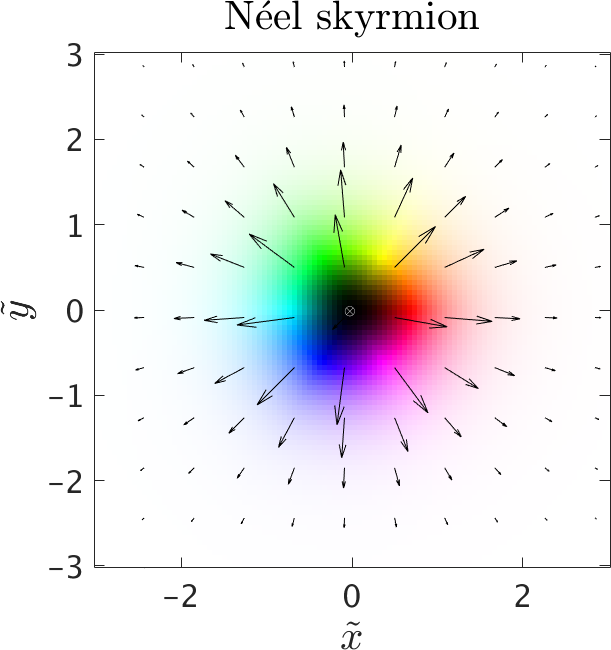}}\ 
  \subfloat[]{\includegraphics[width=0.24\linewidth]{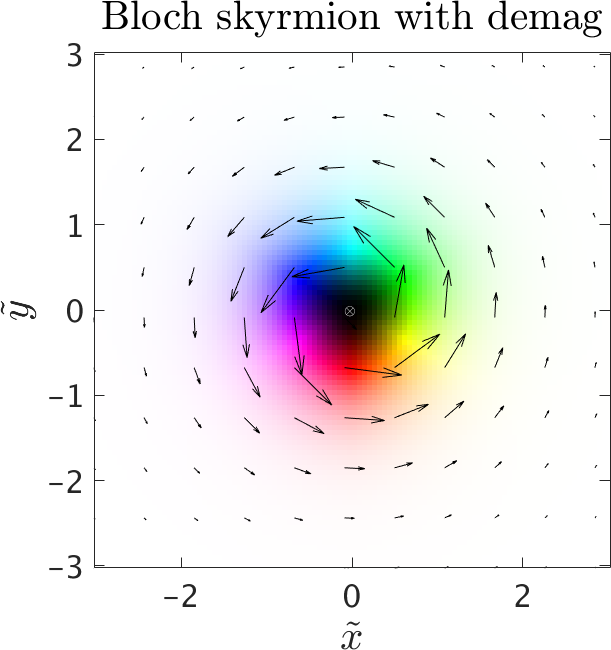}}\ 
    \subfloat[]{\includegraphics[width=0.24\linewidth]{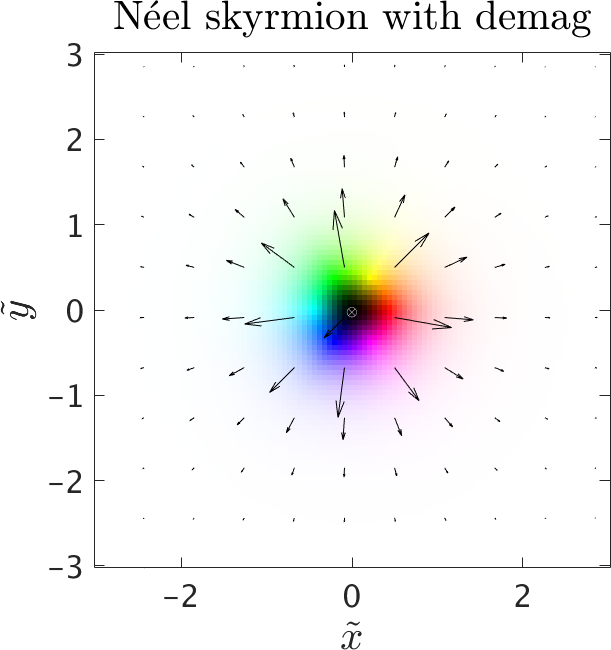}}}
  \caption{(a) Bloch without demag., (b) N\'eel without demag., (c)
    Bloch with demag.~and (d) N\'eel with demag.
  The arrows display the magnetization vector in the plane and are in
  one-to-one correspondence with the coloring of the skyrmions.
  For instance, an arrow pointing in the $\hat{x}$-direction
  corresponds to red.
  White and black correspond to $n_3=+1$ and $n_3=-1$, respectively.
  From now on, we will display the skyrmions using only the coloring,
  which has the same information as the arrows.
  }
  \label{fig:Bloch_Neel_coloring}
\end{figure}
Numerical solutions to the magnetic skyrmion profile functions are
shown in Fig.~\ref{fig:Neel_demag}(a) with the dashed lines
corresponding to the skyrmions without demagnetization taken into
account and the solid lines corresponding to the N\'eel-type skyrmion
with the demagnetization taken into account.
The different colors corresponds to different values of the DMI
coupling $\kappa$.
Since the Bloch- and the N\'eel-type skyrmions have the same profile
functions without the demagnetization taken into account, those
profiles are for both the Bloch and N\'eel skyrmions and also for the
Bloch skyrmions with demagnetization taken into account, since it does
not affect the isolated Bloch skyrmions.
Fig.~\ref{fig:Neel_demag}(b) shows the energy density,
Fig.~\ref{fig:Neel_demag}(c) shows the topological charge density, and
Fig.~\ref{fig:Neel_demag}(d) shows the demagnetization energy, with the curves corresponding to the profiles of
Fig.~\ref{fig:Neel_demag}(a).
In addition to the profile function $f(r)$ of the magnetic skyrmion,
we show the full numerical solution with the magnetization vector
$\bn$ illustrated by arrows mapping $(n_1,n_2)\mapsto(\tilde{x},\tilde{y})$ in
Fig.~\ref{fig:Bloch_Neel_coloring}.
The figure is also colored using a map of the skyrmion's 2-sphere
target space to Runge's color sphere, where the colors are in
one-to-one correspondence with the arrows, as can be seen in the
figure.
The transition from black to white corresponds to the value of $n_3$
interpolating from $n_3=+1$ to $n_3=-1$.
Fig.~\ref{fig:Bloch_Neel_coloring}(a) and (c) show the Bloch skyrmion
without and with demagnetization taken into account and indeed they
are identical (these are full numerical computations), whereas
Figs.~\ref{fig:Bloch_Neel_coloring}(b) and (d) show the N\'eel skyrmion
without and with demagnetization taken into account; the latter has
shrunk about 12\% compared with the former, as is clear from the
figure.

We now turn to the N\'eel-type DW, whose equation of
motion is given by 
\beq
f'' - \frac12\sin(2f) - \eta\cos(f)\p_{\tilde{x}}\Phi = 0,
\label{eq:eomfDW_Neel}
\eeq
as well as $\sin\alpha=0$ which again arise from the first variation
of Eq.~\eqref{eq:E} and the above equation for $\eta=0$ is just the
sine-Gordon equation, which also derives from the energy functional
(in dimensionless units)
\begin{align}
  \widetilde{E}^{\rm DW} &=
  \int\left[\frac12(f')^2
    +\frac12\sin^2f
    +\kappa\cos(\alpha)f'
    +\eta\sin f\left(\cos\alpha\p_{\tilde{x}}\Phi + \sin\alpha\p_{\tilde{y}}\Phi\right)
    \right]\d^2\tilde{x},
  \label{eq:EDW_Neel}
\end{align}
and now $f'=f'(\tilde{x})$.
$\alpha=0$ minimizes the DMI energy, which leaves only one coupling
between $f$ and the demagnetization field: $\p_{\tilde{x}}\Phi$.
The Poisson equation remains that of Eq.~\eqref{eq:Poisson_BlochDW}
and hence sources the magnetic scalar potential for $\alpha=0$.
Integrating the Poisson equation \eqref{eq:Poisson_BlochDW} for
$\alpha=0$, we get
\beq
\p_{\tilde{x}}\Phi = \sin f.
\label{eq:Poisson_1st_order}
\eeq
Although to determine the magnetic scalar potential $\Phi$ one needs
one more integration, the equation of motion for $f$ depends only on
$\p_{\tilde{x}}\Phi$, which is determined in the above equation.
We thus arrive at
\beq
f'' - \frac{1+\eta}{2}\sin(2f),
\eeq
which is again solved by the sine-Gordon solution, albeit with a mass
different from unity:
\beq
f = 2\arctan\left(e^{-\sqrt{1+\eta}(\tilde{x}-X_0)}\right),
\eeq
which corresponds to the field in Riemann sphere coordinates
\beq
u^{\rm DW} = e^{\i\alpha - \sqrt{1+\eta}(\tilde{x}-X_0)}.
\eeq
The DW is thus thinner and for $\eta=0.3$ it is about 12\% thinner
than without the demagnetization field taken into account. 
From this solution, Eq.~\eqref{eq:Poisson_1st_order} can be integrated
to obtain the magnetic scalar potential
\beq
\Phi(\tilde{x}) = 2\int_{-\infty}^{\tilde{x}}\arctan\left(e^{-\sqrt{1+\eta}(\xi-X_0)}\right)\d\xi,
\eeq
which can easily be done numerically. 
Integrating the energy of the N\'eel DW with backreacted
demagnetization field, we get
\beq
\widetilde{E}^{\rm DW} =
Y\left[\frac{2+\eta}{\sqrt{1+\eta}}
  -\cos\alpha\left(\pi\kappa - \frac{2\eta}{\sqrt{1+\eta}}\right)
  \right],
\eeq
which for $\eta:=0$ is minimized by $\alpha=0$.
Interestingly, the N\'eel-type DW will switch ground state to
$\alpha=\pi/2$ if
\beq
\kappa < \kappa^{\rm crit} = \frac{2\eta}{\pi\sqrt{1+\eta}}.
\eeq
The numbers are not that far from each other, since we use physical
parameters for which $\eta=0.3$ that corresponds to
$\kappa^{\rm crit}\approx0.17$, whereas we have chosen $\kappa=0.4$.
The ground state thus remains in the $\alpha=0$ state for these
parameters, but a smaller $\kappa$ could make the situation where the
demagnetization field alters the ground state possible.

\section{Initial conditions}\label{sec:initial_cond}

We will use the same setup as in
Ref.~\cite{Gudnason:2024shv} with the magnetic skyrmion placed at the
origin and the DW placed at $\tilde{x}=X_0<0$, see
Fig.~\ref{fig:setup}. 
\begin{figure}[!htp]
\begin{center}
\begin{tikzpicture}[scale=0.35]
  \draw [<->,very thick] (8.5,9) -- (8.5,8) -- (9.5,8);
  \draw (7.75,8.75) node {{\large $\tilde{y}$}};
  \draw (9.5,7.25) node {{\large $\tilde{x}$}};
  \draw (-10,-10) rectangle (10,10);
  \filldraw [gray] (-5.25,-10) rectangle (-4.75,10);
  \draw (-5.25,-10) -- (-5.25,10);
  \draw (-4.75,-10) -- (-4.75,10);
  \filldraw [gray] (0,0) circle (0.46);
  \draw [<->,very thick] (-4.75,0) -- (-0.46,0);
  \draw (-2.5,1) node {\large $|X_0|$};
  \draw (1,-1.25) node {\large $(0,0)$};
\end{tikzpicture}
\caption{Setup of DW and isolated skyrmion as initial condition.
This figure is taken from Ref.~\cite{Gudnason:2024shv}. }
\label{fig:setup}
\end{center}
\end{figure}
The DW is a solution to the equation of motion \eqref{eq:eom} without
the skyrmion and the skyrmion at the origin is a solution to the
equation of motion without the DW.
They are then superposed in stereographic coordinates as
\beq
u^{\rm composite} = u^{\rm sk} + u^{\rm DW},
\label{eq:u_composite}
\eeq
where the constituent solitons have been discussed for both Bloch-type
and N\'eel-type DMIs, with and without demagnetization, in the
previous section.
In all cases, the superposed field $u^{\rm composite}$ is no longer a
solution to the equation of motion \eqref{eq:eom}, but serves as the
initial condition for our numerical computations.
The advantage of superposing solutions according to
Eq.~\eqref{eq:u_composite} is that the nonlinear sigma model
constraint $\bn\cdot\bn=1$ is automatically satisfied.

If we do not take into account the demagnetization field, the
Bloch-type and N\'eel-type DMIs give rise to physically the same
solutions with the same flows under the LLG equation.
So in this case, we shall just use the Bloch-type DMI and
$\beta=\pi/2$ which corresponds to the skyrmion ground state.
When considering the demagnetization field, $\beta=\pi/2$ for
the Bloch-type DMI and $\beta=0$ for N\'eel-type DMI.

The situation for the DW is similar: its ground state for Bloch-type
DMI is $\alpha=\pi/2$ and for N\'eel-type DMI it is $\alpha=0$.
We cannot freely choose $\alpha$ to be different from
its ground-state value, without physically altering the system as the
initial condition.
This is necessary, since we have shown in Ref.~\cite{Gudnason:2024shv}
that the magnetic skyrmion and the DW repel each other in their
respective ground states -- this is true independently of whether a
Bloch-type or N\'eel-type DMI is considered.

\begin{figure}[!htp]
  \centering
  \includegraphics[width=0.5\linewidth]{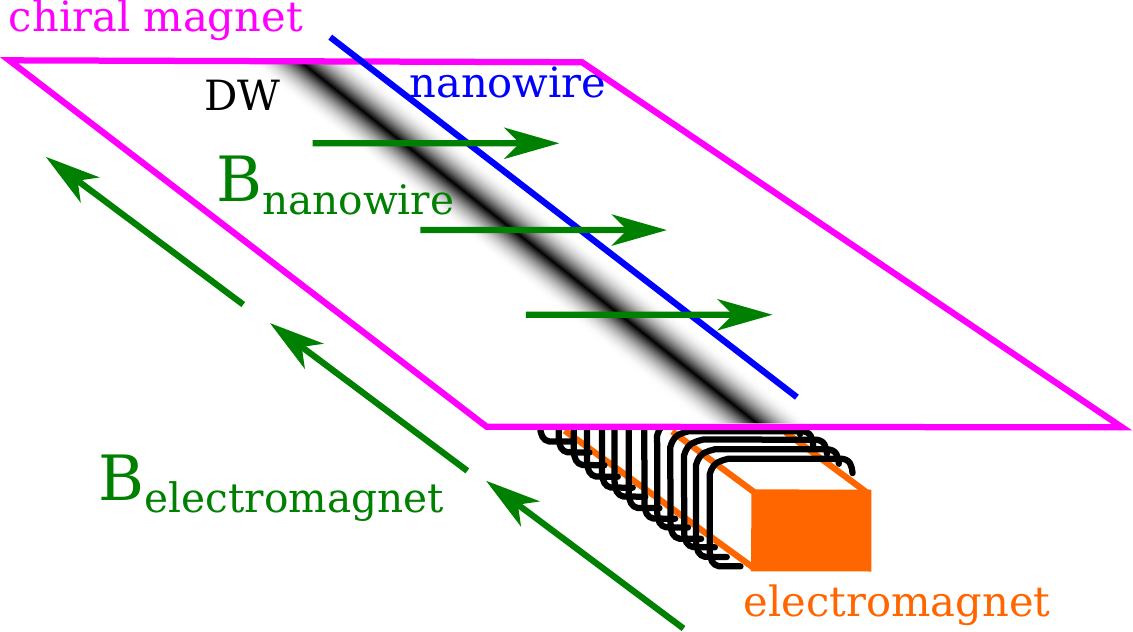}
  \caption{Sketch of a setup that could give rise to the magnetic
    fields described in Eq.~\eqref{eq:extpot}. }
  \label{fig:extmagnets}
\end{figure}
We may, however, physically alter the magnetic system using
electromagnets and (nano)wires to create a magnetic field inducing a
Zeeman term as the initial condition.
The idea is that we switch on such a Zeeman term long before the
experiment starts at time $t=0$ and at time $t=0$, we switch off the
Zeeman term.
We want to localize the external magnetic field very close to the DW
in order to alter its ``modulus'' $\alpha$ away from its ground-state
value, determined by minimizing the DMI energy. 
The exact design, extent and distortion of the actual physical
magnetic field is beyond the scope of this paper.
We trust our friends in the engineering department can create a
suitable design and physical device for an actual experiment.
Schematically, we envision an external potential of the form (in
dimensionless units)
\beq
\widetilde{E}^{\rm local-Zeeman} = \int\left[
\widetilde{B}_{\rm electromagnet}h(\tilde{x}-X_0)(1+\be_2\cdot\bn)
+\widetilde{B}_{\rm nanowire}g(\tilde{x}-X_0)(1+\be_1\cdot\bn)
\right]\d^2\tilde{x},
\label{eq:extpot}
\eeq
where $h$ and $g$ may be thought of as Dirac delta functions for
mathematical localization and more physically spread out distributions
for realistic designs, see Fig.~\ref{fig:extmagnets}.
The ground state (vacuum) with the above Zeeman terms added to the
energy functional \eqref{eq:E} is given by
\beq
\bn=\begin{cases}
\big(-\rho\cos\varphi,-\rho\sin\varphi,\sqrt{1-\rho^2}\big),& 0<\rho<1,\\
\big(-\cos\varphi,-\sin\varphi,0\big),&\rho\geq1,
\end{cases}
\eeq
where
\beq
\rho := \sqrt{\widetilde{B}_{\rm electromagnet}^2+\widetilde{B}_{\rm nanowire}^2},\qquad
\varphi=\arctan\left(\frac{\widetilde{B}_{\rm electromagnet}}{\widetilde{B}_{\rm nanowire}}\right).
\eeq
In order to dominate over the DMI that also wants to align the
magnetization inside the DW, we need to take $\rho>\kappa$.
Then the direction of the magnetization in the plane is given by
$\varphi$.
Comparing now with the DW field's phase $\alpha$, we see that
$\alpha=\varphi$ in the temporary ground state created by the
electromagnet and the nanowire.
In the rest of this paper, we shall simply refer to the initial
condition of the DW as $\alpha$ at time $t=0$ or $\alpha_0$.

\section{Numerical method}\label{sec:num_method}

We will evolve the LLG flow using the fourth-order Runge-Kutta (RK4)
method and calculate the discrete spatial derivatives using a
finite-difference method to fourth order in the discretization on a
5-point equidistant lattice.
That is, the time derivative of the LLG equation \eqref{eq:LLG}
is evolved with RK4 and the spatial derivatives of Eq.~\eqref{eq:eom}
are evaluated on a lattice with the finite-difference method.
The lattice size is set to be $682^2$ with spatial stepsize
$h_{\tilde{x}}=h_{\tilde{y}}=0.0587$ and the temporal stepsize
$h_{\tilde{t}}=6\times10^{-4}$ -- all in dimensionless units.

The demagnetization effect is taken into account by computing the
magnetic scalar potential $\Phi$ at every (full) step of the RK4 flow,
using the conjugate-gradients method -- that is, the Poisson equation
\eqref{eq:Poisson} is solved at every step of the LLG flow, so that
$\Phi$ is always a solution even though $\bn$ is not (necessarily).
This method was originally developed in high-energy physics
\cite{Gudnason:2020arj,Harland:2024dca} and has also recently been
used in condensed matter physics \cite{Leask:2025pdz}.  
Physically, this could be interpreted as the electromagnetic field
propagating at a speed faster than the change in the magnetization of
the material in question and avoids the caveat of instabilities 
related to the propagation of gauge fields.

The initial condition for the computations is given in
Sec.~\ref{sec:initial_cond} and the boundary conditions (BCs) for the
magnetization vector $\bn$ is Dirichlet on the left- and right-hand
sides of the lattice and Neumann on the top and bottom:
$\bn(-X/2,\tilde{y})=(0,0,-1)$, $\bn(X/2,\tilde{y})=(0,0,1)$,
$\p_{\tilde{y}}\bn(\tilde{x},-Y/2)=\p_{\tilde{y}}\bn(\tilde{x},Y/2)=0$ with $X$ and $Y$
being the width and height of the lattice in dimensionless units,
respectively.
The Dirichlet BCs are natural to enforce the two opposite ground
states (vacua) on each side of the DW and the Neumann BCs are
compatible with the ground state of the DW.
It does, however, allow the magnetic skyrmion in the bulk to exit the
simulation domain, but that does never happen in the computation we
are doing, so this is irrelevant.
It does, however, also allow the bound states -- the DW-skyrmions --
to exit at the top and the bottom of the simulation domain.
For a discussion of this issue, see the discussion below.

\section{Numerical results}\label{sec:results}

In Ref.~\cite{Gudnason:2024shv}, we have shown by analytic methods
that at asymptotic distances, the magnetic skyrmion (in the bulk of
the material) and the DW repel each other, when they are both in their
ground states -- this is the reason for introducing perturbations in
the initial condition in Sec.~\ref{sec:initial_cond}.
We also demonstrated in Ref.~\cite{Gudnason:2024shv} that the magnetic
skyrmion in close proximity to the DW can be destroyed.
This happens when the skyrmion is close, but not close enough to the
DW to be captured and transformed into a DW-skyrmion.
Physically, we can understand this in the following way.
The magnetic skyrmion is repelled by the DW and has a negative DMI
energy.
If we can attract the skyrmion into the DW, it will turn its DMI
energy positive and be stabilized as a 1-dimensional soliton living on
the host DW.
If the skyrmion fails to enter properly the DW, but is so close to the
DW that its DMI energy is either positive or nearly vanishing, it
risks a collapse.
The Derrick's theorem scaling argument that stabilizes the magnetic
skyrmion fails when the DMI energy is not negative. 
All the computations done in Ref.~\cite{Gudnason:2024shv} were done by
considering only energy minimization -- a method that is about 2
orders of magnitude faster and computationally cheaper than using the
LLG flow.
In this paper, we use the LLG flow as it is more physical for
dynamical questions and does show slight deviations from the fast
energy minimization techniques used in Ref.~\cite{Gudnason:2024shv}.

\subsection{Final states}

We will now describe the final states, that are obtained by evolving
the LLG flow to a stationary point within the setup described in the
previous section.
We should point out that the minute details of which final states
appear and which do not, depend on the size of the magnetic material
(or in our case, the size of the simulation box) as well as on the
boundary conditions.
This means that the results obtained at the unstable
point of the DW (i.e.~$\alpha=\tfrac{3\pi}{2}$ for the Bloch DMI and
$\alpha=\pi$ for the N\'eel DMI), are very much dependent on such
details -- different sizes of the material will give different
results, as can also be understood from the videos in the ancillary files; see also the topological charge as a function of
  time, $Q(\tilde{t})$, in Appendix \ref{app:topo_charge}.

\begin{figure}[!htp]
  \centering
  \includegraphics[width=0.8\linewidth]{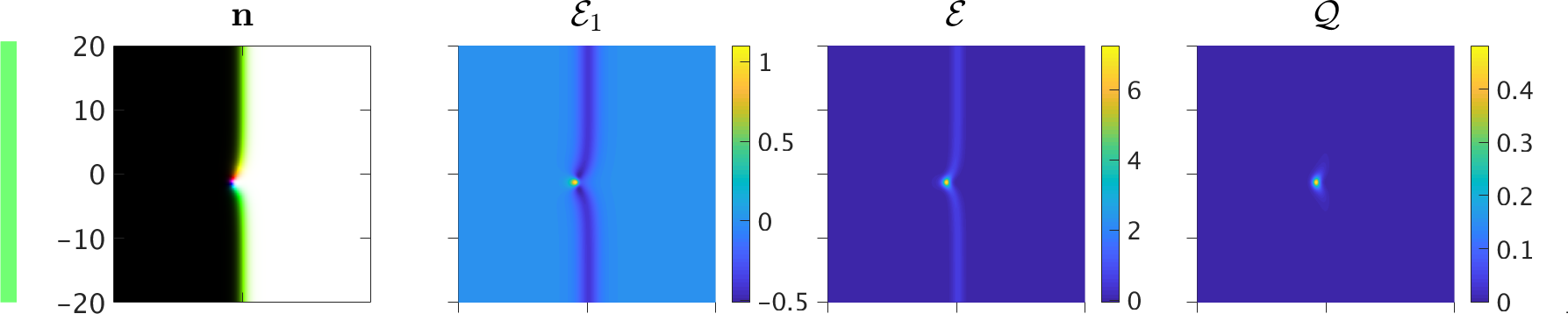}
  \includegraphics[width=0.8\linewidth]{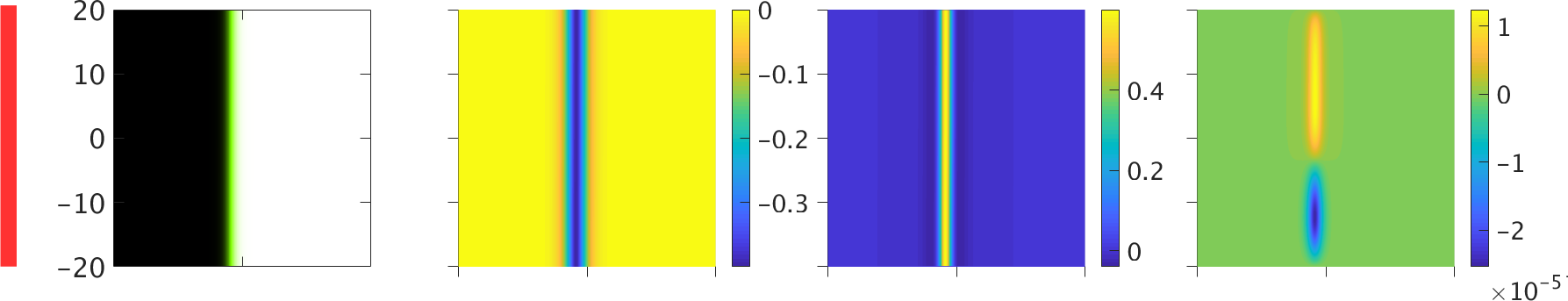}
  \includegraphics[width=0.8\linewidth]{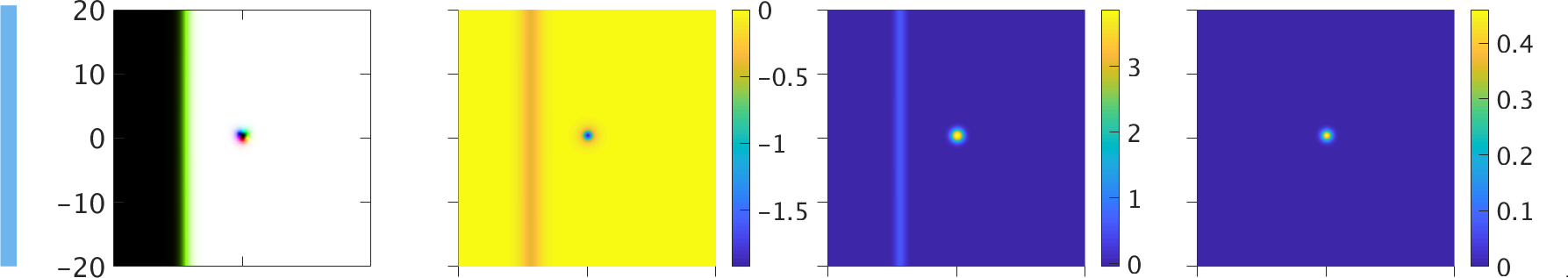}
  \includegraphics[width=0.8\linewidth]{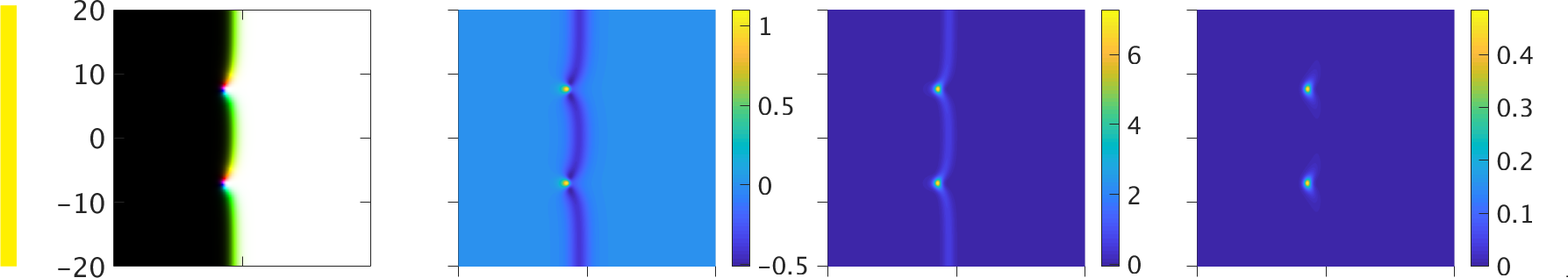}
  \includegraphics[width=0.8\linewidth]{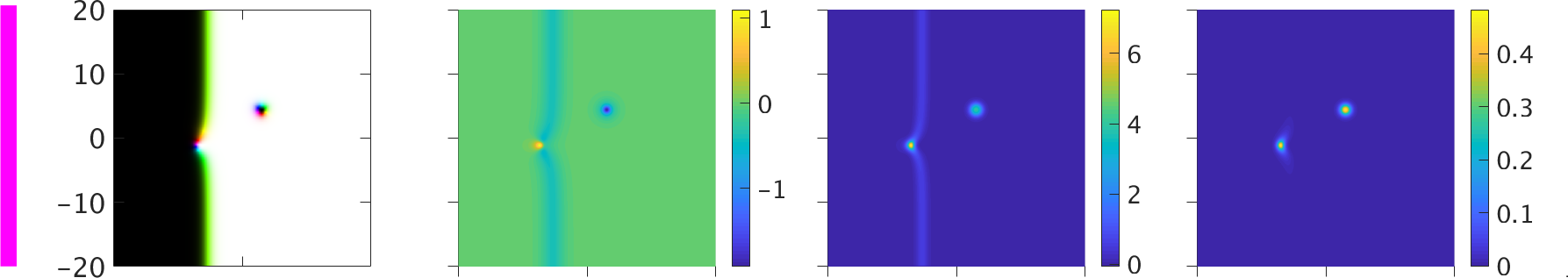}
  \includegraphics[width=0.8\linewidth]{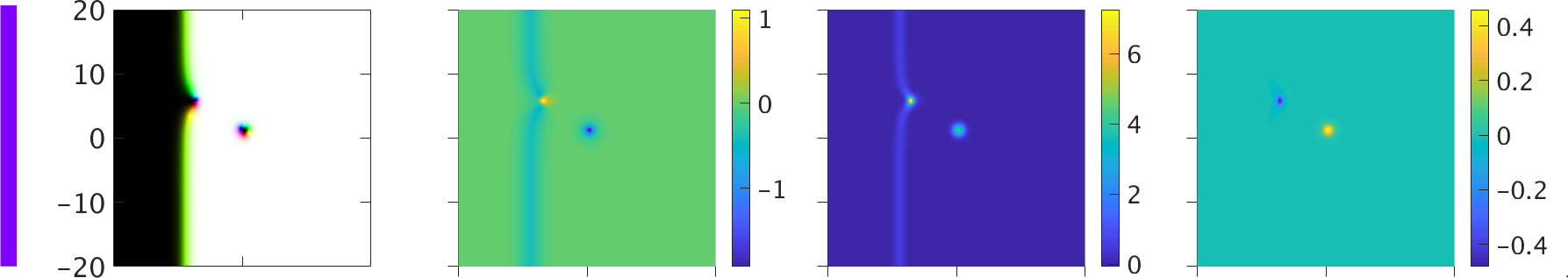}
  \includegraphics[width=0.8\linewidth]{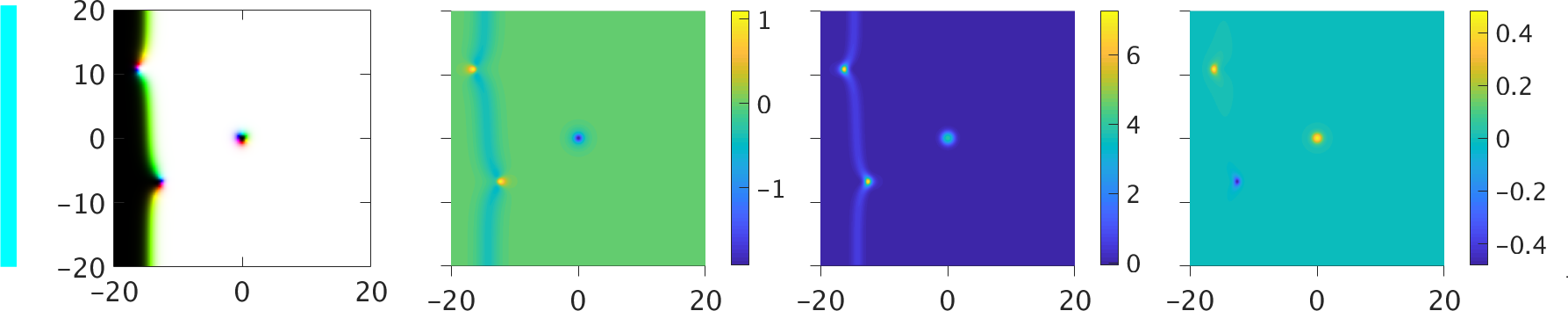}
  \caption{Final states of evolution of the LLG equation from the initial
    condition \eqref{eq:u_composite} in the case of Bloch DMI without
    demagnetization (and equivalently N\'eel DMI without
    demagnetization by the map from
    Fig.~\ref{fig:Bloch_Neel_coloring}(a) to
    Fig.~\ref{fig:Bloch_Neel_coloring}(b)).
    The columns display the color code for the final state
    (which is referred to in the phase diagram in
      Fig.~\ref{fig:phasediagram} and should not be confused with the
      colors representing the magnetization vector as in Fig.~\ref{fig:Bloch_Neel_coloring}), the
    magnetization vector (for a map to vectors, see
    Fig.~\ref{fig:Bloch_Neel_coloring}), the DMI energy density, the total
    energy density and finally the topological charge density.
    The rows correspond to a DW-skyrmion (green/C), an empty DW
    (red/A), a bulk skyrmion (blue/B), two DW-skyrmions
    (yellow/D), a DW-skyrmion and a bulk skyrmion (magenta/F), an
    anti-DW-skyrmion and a bulk skyrmion (purple/G) and finally a
    DW-skyrmion-anti-DW-skyrmion pair with a bulk skyrmion (cyan/H).
    The magnetization vector for the DW with Bloch DMI
    interpolates, from left to right, as
    $-\hat{z}\to\hat{y}\to\hat{z}$ shown with colors as black $\to$
    green $\to$ white.
  }
  \label{fig:finalstates}
\end{figure}

\begin{figure}[!htp]
  \centering
  \includegraphics[width=\linewidth]{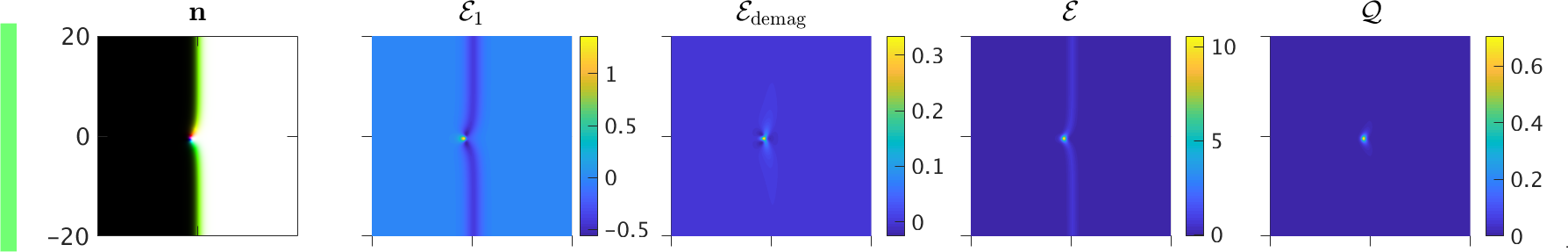}
  \includegraphics[width=\linewidth]{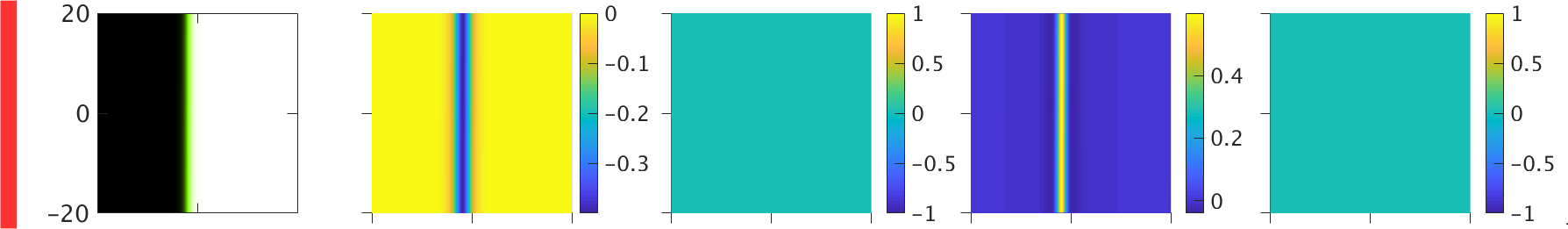}
  \includegraphics[width=\linewidth]{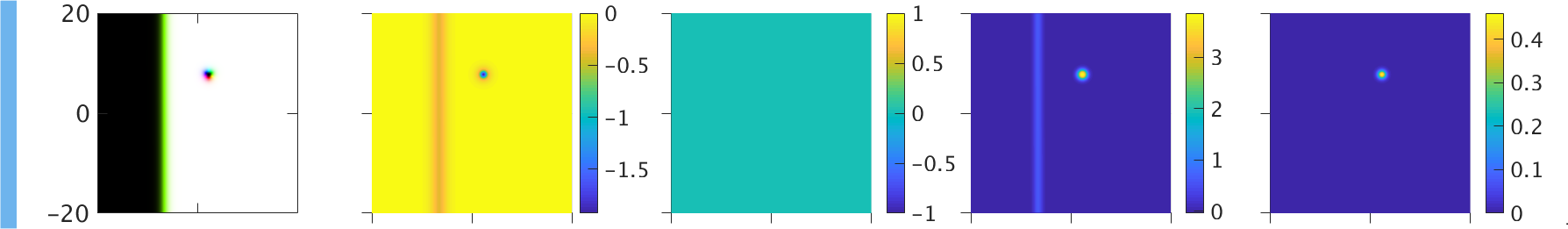}
  \includegraphics[width=\linewidth]{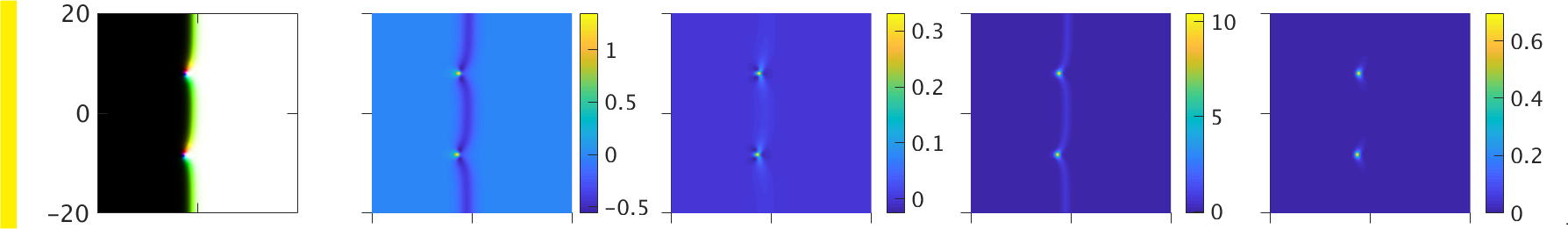}
  \includegraphics[width=\linewidth]{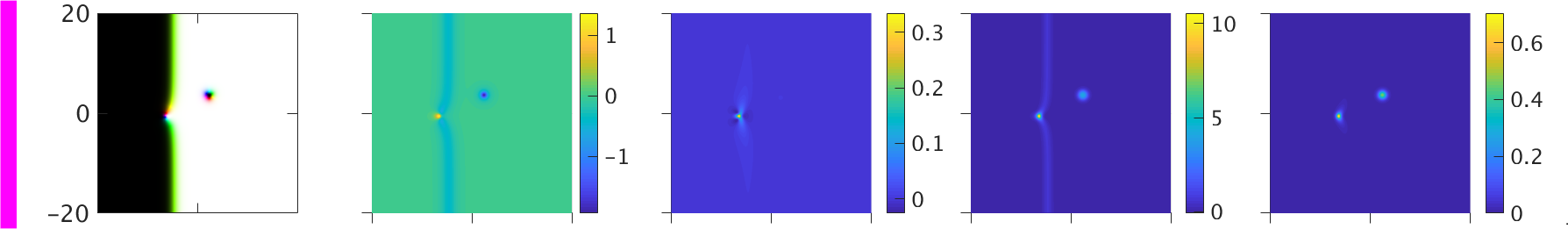}
  \includegraphics[width=\linewidth]{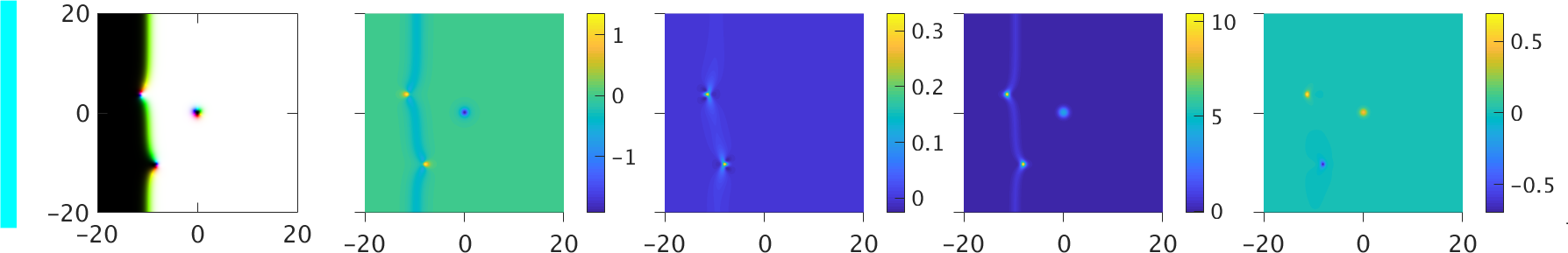}
  \caption{Final states of evolution of the LLG equation from the initial
    condition \eqref{eq:u_composite} in the case of Bloch DMI with
    demagnetization.
    The columns display the color code for the final state (which is referred to in the phase diagram in Fig.~\ref{fig:phasediagram_Bdm} and should not be confused with the
      colors representing the magnetization vector as in Fig.~\ref{fig:Bloch_Neel_coloring}), the
    magnetization vector (for a map to vectors, see
    Fig.~\ref{fig:Bloch_Neel_coloring}), the DMI energy density, the
    demagnetization energy density, the total energy density and
    finally the topological charge density.
    The rows correspond to a DW-skyrmion (green/C), an empty DW
    (red/A), a bulk skyrmion (blue/B), two DW-skyrmions
    (yellow/D), a DW-skyrmion and a bulk skyrmion (magenta/F) and
    finally a DW-skyrmion-anti-DW-skyrmion pair with a bulk skyrmion
    (cyan/H).
    The magnetization vector for the DW with Bloch DMI
    interpolates, from left to right, as
    $-\hat{z}\to\hat{y}\to\hat{z}$ shown with colors as black $\to$
    green $\to$ white.
  }
  \label{fig:finalstates_Bdm}
\end{figure}

\begin{figure}[!htp]
  \centering
  \includegraphics[width=\linewidth]{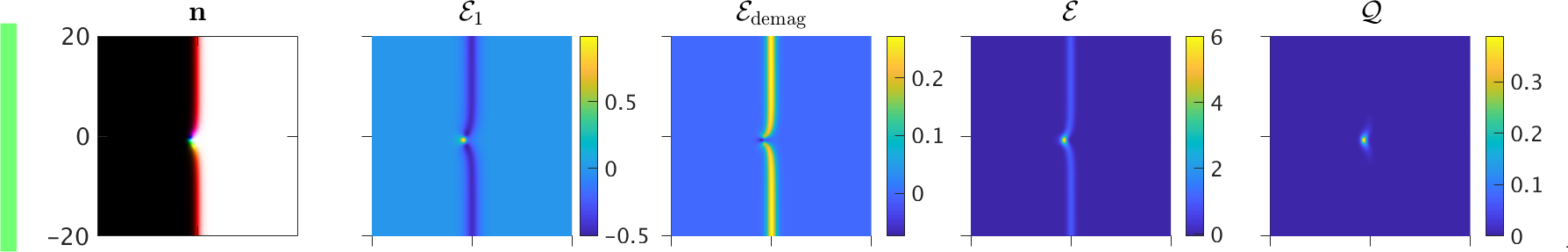}
  \includegraphics[width=\linewidth]{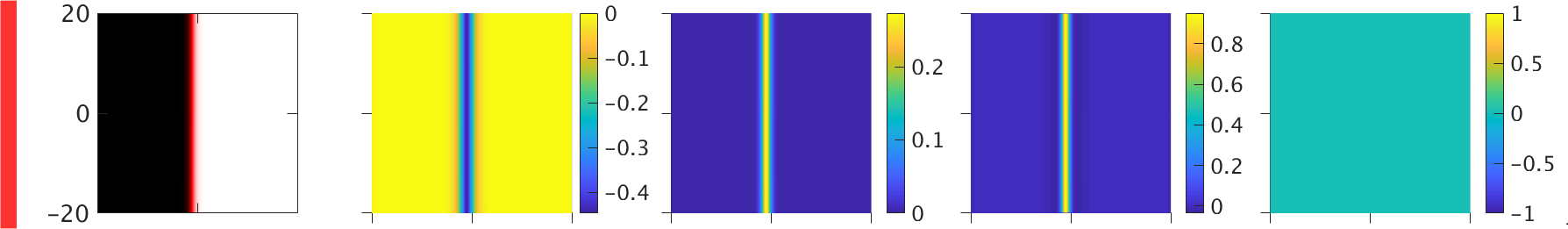}
  \includegraphics[width=\linewidth]{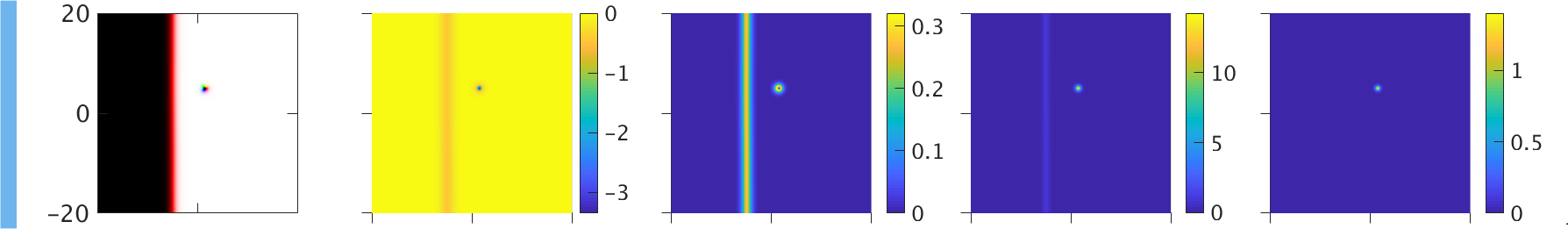}
  \includegraphics[width=\linewidth]{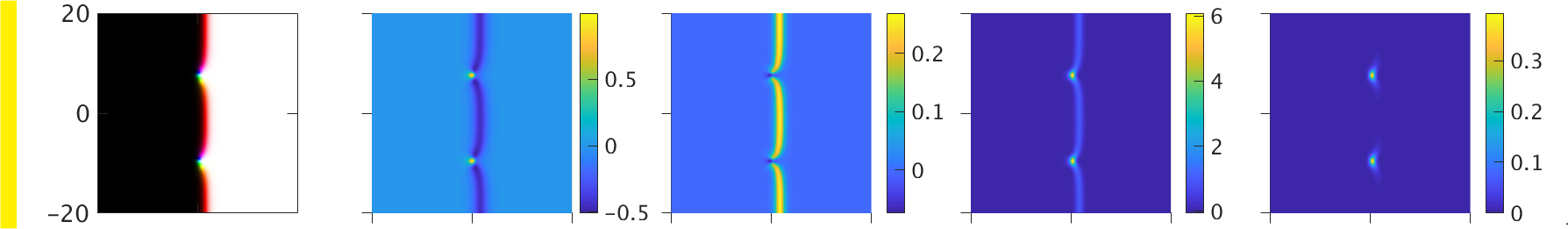}
  \includegraphics[width=\linewidth]{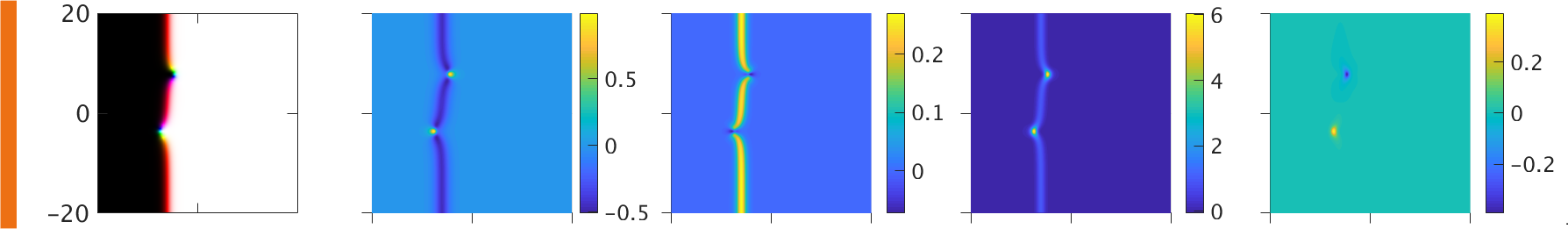}
  \includegraphics[width=\linewidth]{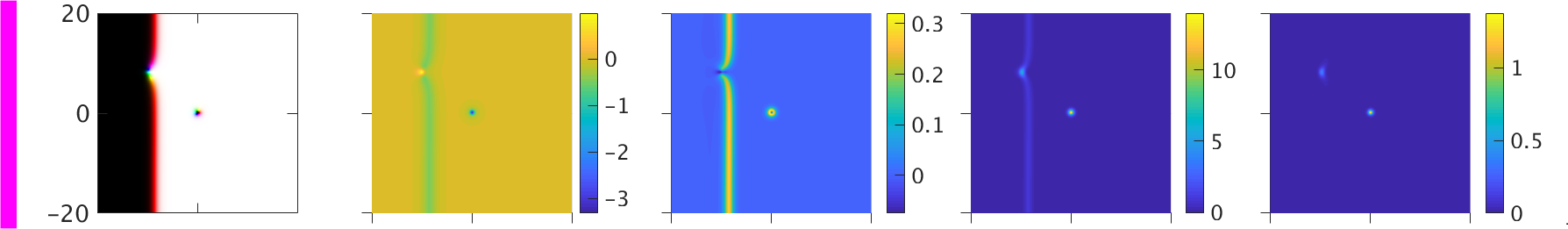}
  \includegraphics[width=\linewidth]{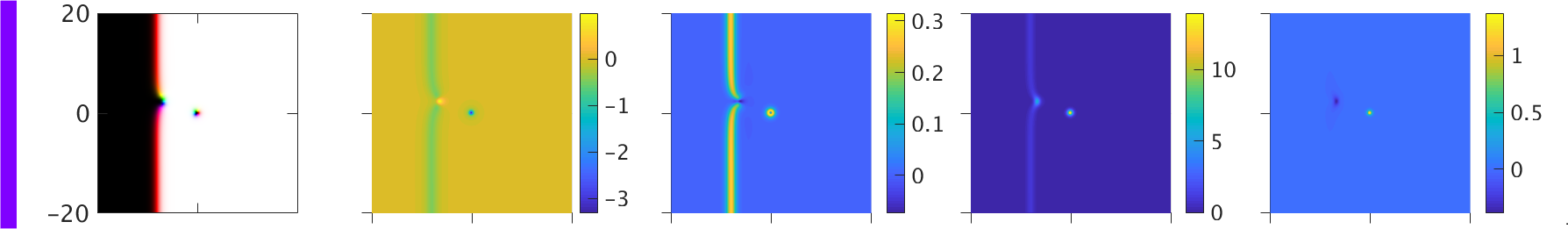}
  \includegraphics[width=\linewidth]{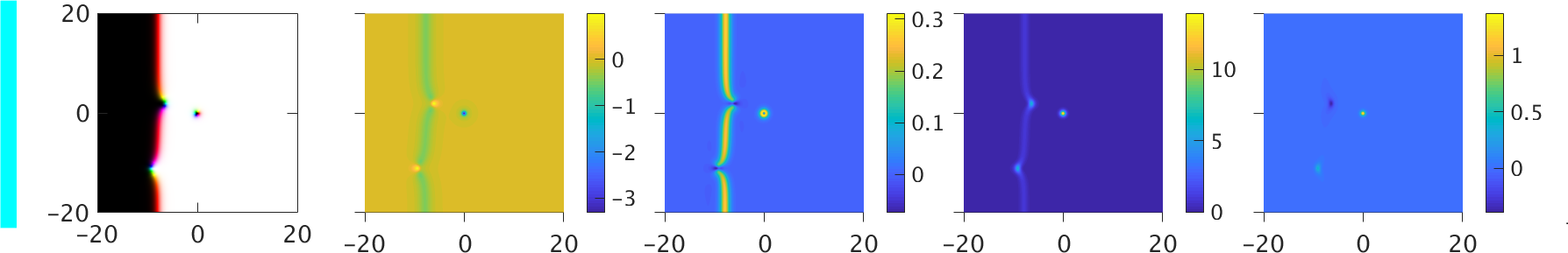}
  \caption{Final states of evolution of the LLG equation from the initial
    condition \eqref{eq:u_composite} in the case of N\'eel DMI with
    demagnetization.
    The columns display the color code for the final state (which is referred to in the phase diagram in Fig.~\ref{fig:phasediagram_Ndm} and should not be confused with the
      colors representing the magnetization vector as in Fig.~\ref{fig:Bloch_Neel_coloring}), the
    magnetization vector (for a map to vectors, see
    Fig.~\ref{fig:Bloch_Neel_coloring}), the DMI energy density, the
    demagnetization energy density, the total energy density and
    finally the topological charge density.
    The rows correspond to a DW-skyrmion (green/C), an empty DW
    (red/A), a bulk skyrmion (blue/B), two DW-skyrmions
    (yellow/D), a DW-skyrmion-anti-DW-skyrmion pair (orange/E), a
    DW-skyrmion and a bulk skyrmion (magenta/F), an anti-DW-skyrmion
    and a bulk skyrmion (purple/G) and finally a
    DW-skyrmion-anti-DW-skyrmion pair with a bulk skyrmion (cyan/H).
    The magnetization vector for the DW with N\'eel DMI
    interpolates, from left to right, as
    $-\hat{z}\to\hat{x}\to\hat{z}$ shown with colors as black $\to$ 
    red $\to$ white.
  }
  \label{fig:finalstates_Ndm}
\end{figure}

Figs.~\ref{fig:finalstates} and \ref{fig:finalstates_Bdm} show the
final states for the case of the Bloch DMI without and with
demagnetization, respectively.
By chance, there is one final state less in the case with
demagnetization than without, but no underlying physical reason for
this -- the outcome of the LLG flow from the unstable DW is most
likely chaotic.
The total energies of the magenta and of the purple final state are
identical to within numerical precision -- that is, the DW-skyrmion
and the anti-DW-skyrmion are degenerate in energies.
The color codes are referenced in Figs.~\ref{fig:phasediagram} and
\ref{fig:phasediagram_Bdm} below.

Fig.~\ref{fig:finalstates} also describes the case of N\'eel-type DMI
without demagnetization with a simple map that rotates the
magnetization vectors as illustrated in
Fig.~\ref{fig:Bloch_Neel_coloring}(a) and
Fig.~\ref{fig:Bloch_Neel_coloring}(b).

Finally, Fig.~\ref{fig:finalstates_Ndm} shows the case of N\'eel-type
DMI with demagnetization taken into account.
In this case, there is accidentally a new final state compared to the
case without demagnetization, namely the orange color code
corresponding to a DW-skyrmion-anti-DW-skyrmion pair with a vanishing
total topological charge.
The energy of the double DW-skyrmion is slightly larger than that of
the DW-skyrmion-anti-DW-skyrmion pair.

\subsection{Symplectic flow: DW movement}

One of the major drastic effects of turning from a gradient flow or
arrested Newton flow to an LLG flow due to the symplectic component
(i.e.~the first term) in the LLG equation \eqref{eq:LLG}, is the
movement of the perturbed DW.
The Gilbert damping term is mathematically identical to
gradient flow and for sufficiently (unphysically) large Gilbert
damping coefficients ($\alpha_G$) LLG and gradient flows are
equivalent
(the phase diagram without demagnetization taken into account
was studied in Ref.~\cite{Gudnason:2024shv}).
Nevertheless, for a physical value of the Gilbert damping coefficient
(Eq.~\eqref{eq:modelparms}) the symplectic component weighs in --
especially when the solitons are ``far'' from their ground states in
field space.

Let us consider the DW movement from its initial (perturbed) state to
its ground state, starting with the case of the Bloch DMI.
In order to understand what happens to the position of the DW under
the flow of the phase from $\alpha=\frac{3\pi}{2}$ to
$\alpha=\frac{\pi}{2}$, i.e.~the ground state, we promote
$\alpha\to\alpha(\tilde{t})$ and $X_0\to X_0(\tilde{t})$ in
Eq.~\eqref{eq:u_DW_sol}, treating them as pseudo moduli, and integrate
the LLG equation (i.e.~Eq.~\eqref{eq:LLG}) over $\tilde{x}$ to obtain
the effective equations also known as Thiele equations:
\begin{align}
  \p_{\tilde{t}}\alpha &=
  \alpha_G\cos\alpha\left(\frac{4\kappa}{\pi} + \eta\sin\alpha\right),
  \label{eq:alphat_eff}\\
  \p_{\tilde{t}}X_0 &=
  \cos\alpha\left(\frac{\pi\kappa}{2} +
  \frac{\pi\alpha_G\eta}{4}\cos\alpha + \eta\sin\alpha\right),
  \label{eq:X0t_eff}
\end{align}
where we have used the solution $\p_{\tilde{x}}\Phi=\cos\alpha\sin f$
to the Poisson equation \eqref{eq:Poisson}.
The Thiele equation in condensed matter physics is known as the moduli
space approximation in high-energy physics \cite{Manton:1981mp}.
\begin{figure}[!htp]
  \centering
  \mbox{\subfloat[]{\includegraphics[width=0.49\linewidth]{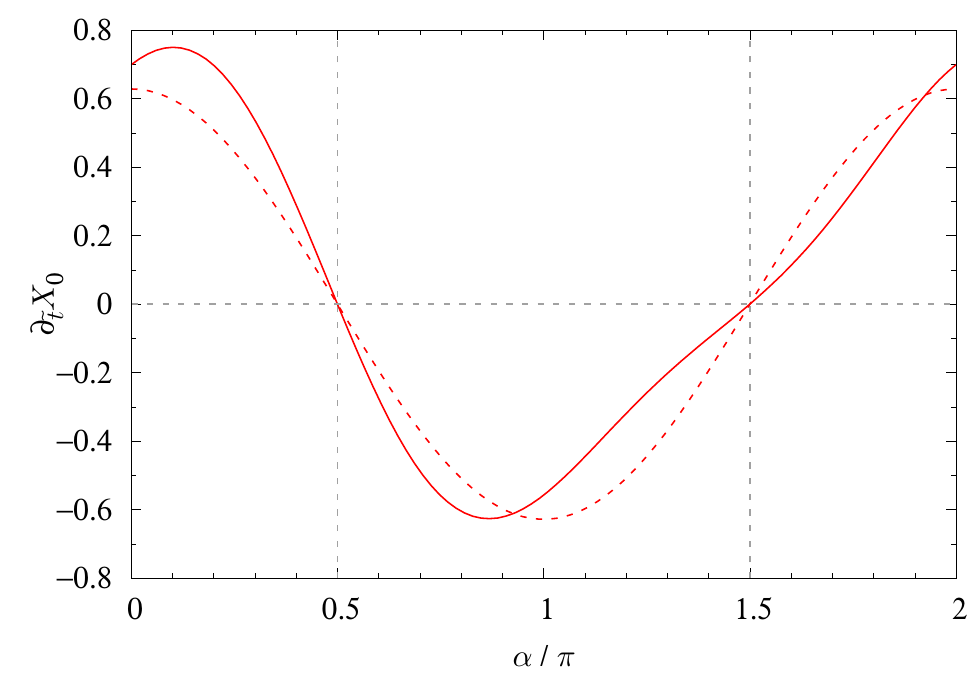}}
    \subfloat[]{\includegraphics[width=0.49\linewidth]{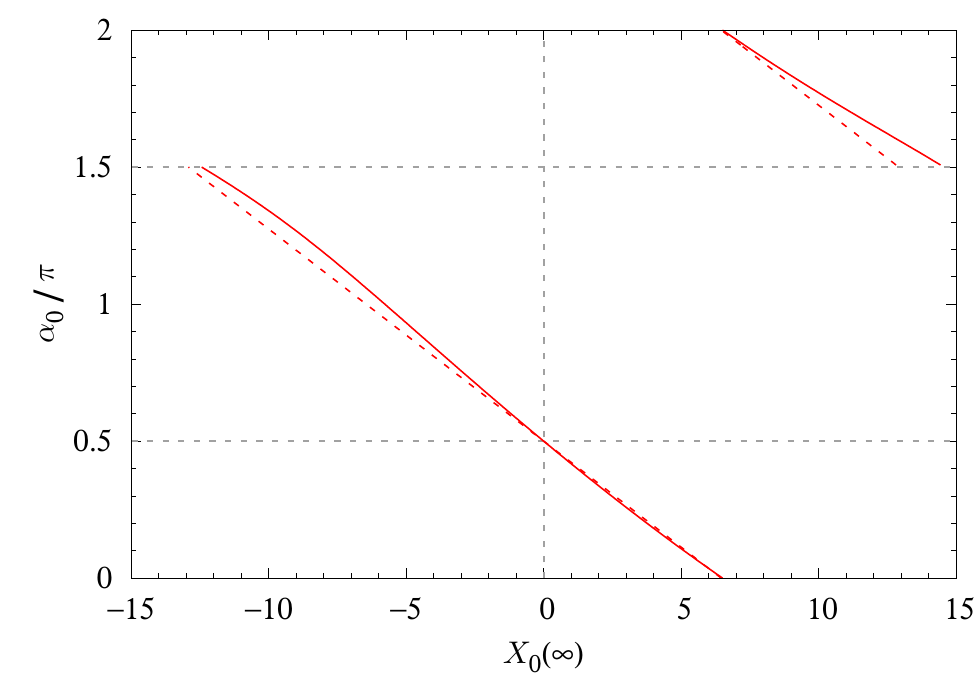}}}
  \mbox{\subfloat[]{\includegraphics[width=0.49\linewidth]{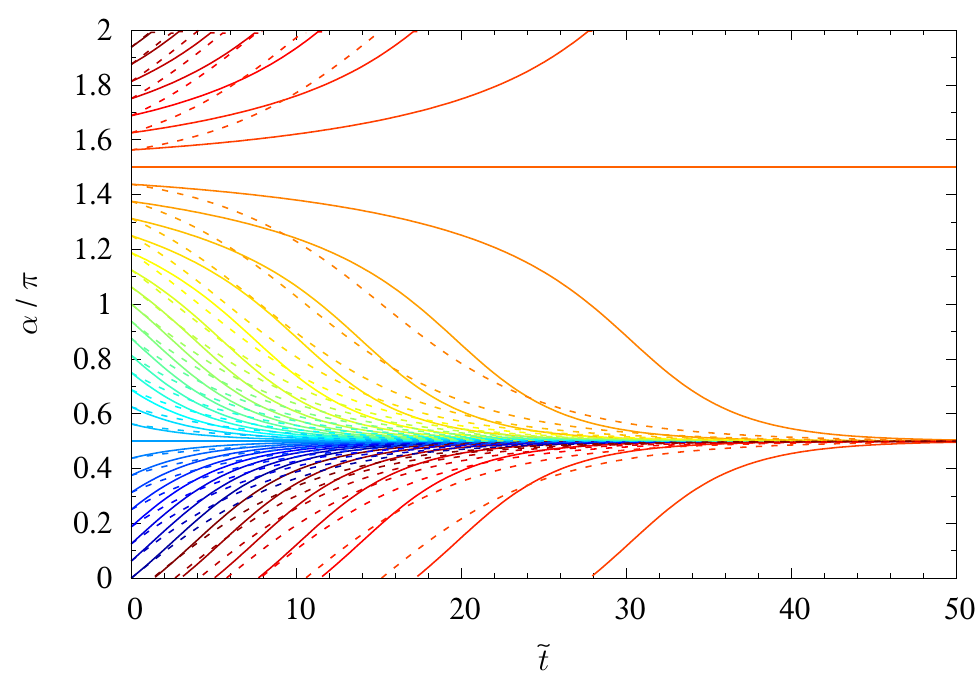}}
    \subfloat[]{\includegraphics[width=0.49\linewidth]{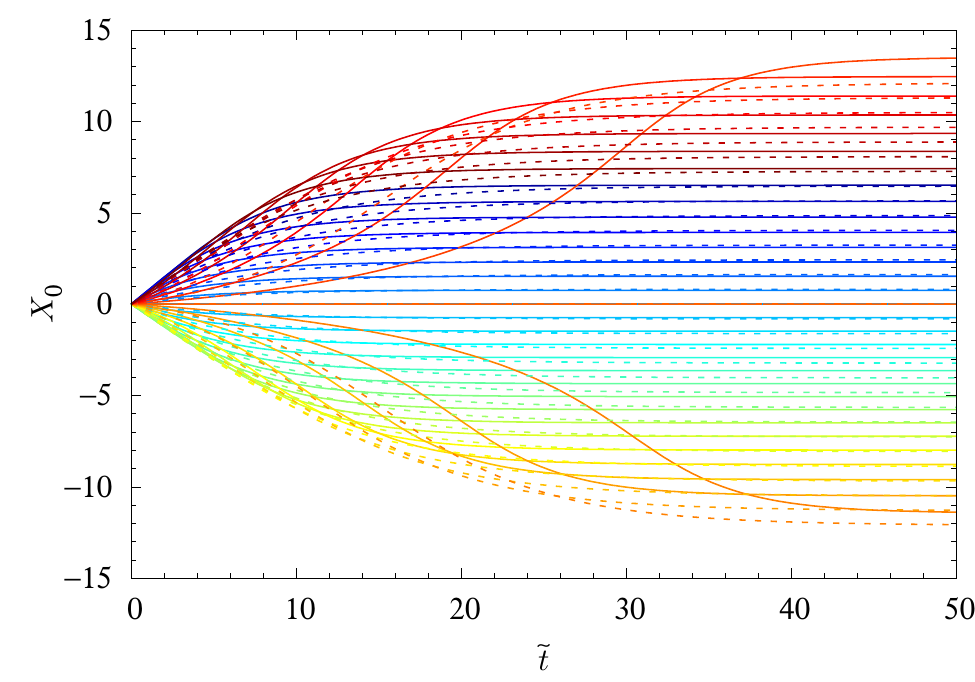}}}
  \caption{Thiele equation dynamics of the DW with a Bloch DMI.
    (a) The Thiele equation \eqref{eq:X0t_eff} as a function of
    the DW phase $\alpha$. (b) The final position of the DW,
    $X_0(\infty)$, as a function of the initial value of the DW phase
    $\alpha_0=\alpha(0)$ (the plot is transposed on purpose for easier
    comparison with the phase diagram).
    (c) The DW angle, $\alpha$, and (d) the DW position, $X_0$, as
    functions of time during Thiele (LLG) flow.
    The color coding between panel (c) and (d) is the same: the
    initial phase $\alpha_0$ in panel (d) corresponds to the left-most
    value of $\alpha$ in panel (c).
    The dashed lines correspond to the exact analytic solutions, whereas
    the solid lines are numerical solutions that take into account the 
    demagnetization effect.
    All quantities in this figure are dimensionless and the values of
    the constants are given in Eq.~\eqref{eq:modelparms}.
  }
  \label{fig:movingDWmodulispace}
\end{figure}
The result is interesting: According to the Thiele equations, the
DW does not flow to its ground state without the Gilbert damping,
although it moves.
The force on the DW is present also without demagnetization and is
driven already purely by the DMI.
For $\alpha\in\left[-\tfrac{\pi}{2},\tfrac{\pi}{2}\right]$ it
moves left and for
$\alpha\in\left[\tfrac{\pi}{2},\tfrac{3\pi}{2}\right]$ it moves right,
which we can see by setting $\eta=\alpha_G=0$ in
Eq.~\eqref{eq:X0t_eff}.
Turning on Gilbert damping ($\alpha_G>0$) ensures that the phase of
the DW will flow to its ground state $\alpha=\pi/2$.
We can also see from Eq.~\eqref{eq:X0t_eff}, that a stable DW (in its
ground state $\alpha=\pi/2$) does not move, but a perturbed DW moves
and the direction is in principle determined by a combination of
factors, see Fig.~\ref{fig:movingDWmodulispace}(a).

The Thiele equation for the DW phase $\alpha$ \eqref{eq:alphat_eff} is
a first-order ordinary differential equation (ODE) and can easily be
solved analytically, without taking into account the effect of
demagnetization
\beq
\alpha(\tilde{t}) =
  2\arctan\left[\tanh\left(\frac{2\alpha_G\kappa\tilde{t}}{\pi}
    +\arctanh\left[\tan\left(\frac{\alpha_0}{2}\right)\right]\right)\right],
\eeq
where $\alpha_0=\alpha(0)$ is the phase of the DW at time
$\tilde{t}=0$.
The formula is time independent for $\alpha_0=\frac{\pi}{2}$ and
$\alpha_0=\frac{3\pi}{2}$, which is simply due to the fact that the
equations of motion are solved (although the latter corresponds to an
unstable fixed point) and hence, there will be no dynamical evolution
of $\alpha$. 
Inserting this solution into Eq.~\eqref{eq:X0t_eff}, we obtain the
distance that the DW has moved as a function of time in dimensionless
units
\beq
X_0(\tilde{t}) = X_0(0) +
\frac{\pi^2}{4\alpha_G}\left(
\arctan\left[\tanh\left(\frac{2\alpha_G\kappa\tilde{t}}{\pi}
  + \arctanh\left[\tan\left(\frac{\alpha_0}{2}\right)\right]
  \right)\right]
-\frac{\alpha_0}{2}
\right),
\eeq
where $X_0(0)$ is the position of the DW at time $\tilde{t}=0$.
The DW phase ($\alpha$) and position ($X_0$) are plotted as functions
of dimensionless time in Figs.~\ref{fig:movingDWmodulispace}(c) and
\ref{fig:movingDWmodulispace}(d), respectively, whereas the final DW
position $X_0(\tilde{t}\to\infty)$ is shown in
Fig.~\ref{fig:movingDWmodulispace}(b). 
The exact analytic solutions for $\alpha$ and $X_0$ are shown 
in Fig.~\ref{fig:movingDWmodulispace} with dashed lines and numerical 
solutions for the case incorporating the effect of demagnetization with 
$\eta=0.3$ are shown with solid lines.

We now turn to the case of the N\'eel DMI, which changes trivially the
equations \eqref{eq:alphat_eff} and \eqref{eq:X0t_eff} (and hence their
solutions) for vanishing demagnetization (i.e.~$\eta=0$) by mapping
$\alpha\mapsto\alpha+\pi/2$.
On the other hand, the terms due to the demagnetization effect remain
the same as in the Bloch case, but under the mentioned mapping, they
transform.
Hence, the N\'eel case is physically different from the Bloch case
with the demagnetization taken into account.
The equivalent of Eqs.~\eqref{eq:alphat_eff} and \eqref{eq:X0t_eff}
in the N\'eel case are
\begin{align}
  \p_{\tilde{t}}\alpha &=
  \alpha_G\sin\alpha\left(-\frac{4\kappa\sqrt{1+\eta}}{\pi} + \eta\cos\alpha\right),
  \label{eq:Neel_alphat_eff}\\
  \p_{\tilde{t}}X_0 &=\sin\alpha\left(
  -\frac{\pi\kappa}{2}
  -\frac{\pi\alpha_G\eta}{4\sqrt{1+\eta}}\sin\alpha
  +\frac{\eta}{\sqrt{1+\eta}}\cos\alpha\right).
  \label{eq:Neel_X0t_eff}
\end{align}
Due to the change in trigonometric functions, the analytic solution to
the above equations in the limit of $\eta\to0$ is slightly different:
\begin{align}
\alpha(\tilde{t}) &=
  2\arctan\left[\exp\left(-\frac{4\alpha_G\kappa}{\pi}\,\tilde{t}\right)
    \tan\left(\frac{\alpha_0}{2}\right)\right],\\
X_0(\tilde{t}) &= X_0(0) +
\frac{\pi^2}{4\alpha_G}\left(
\arctan\left[\exp\left(-\frac{4\alpha_G\kappa}{\pi}\,\tilde{t}\right)
  \tan\left(\frac{\alpha_0}{2}\right)\right]
-\frac{\alpha_0}{2}
\right),
\end{align}  
which is valid for $\alpha_0\in[0,2\pi]\backslash\pi$.
$\alpha_0=\pi$ is stationary and corresponds to the unstable DW in the
N\'eel DMI case -- hence no movement of the DW (unless perturbed
infinitesimally). 

\begin{figure}[!htp]
  \centering
  \mbox{\subfloat[]{\includegraphics[width=0.49\linewidth]{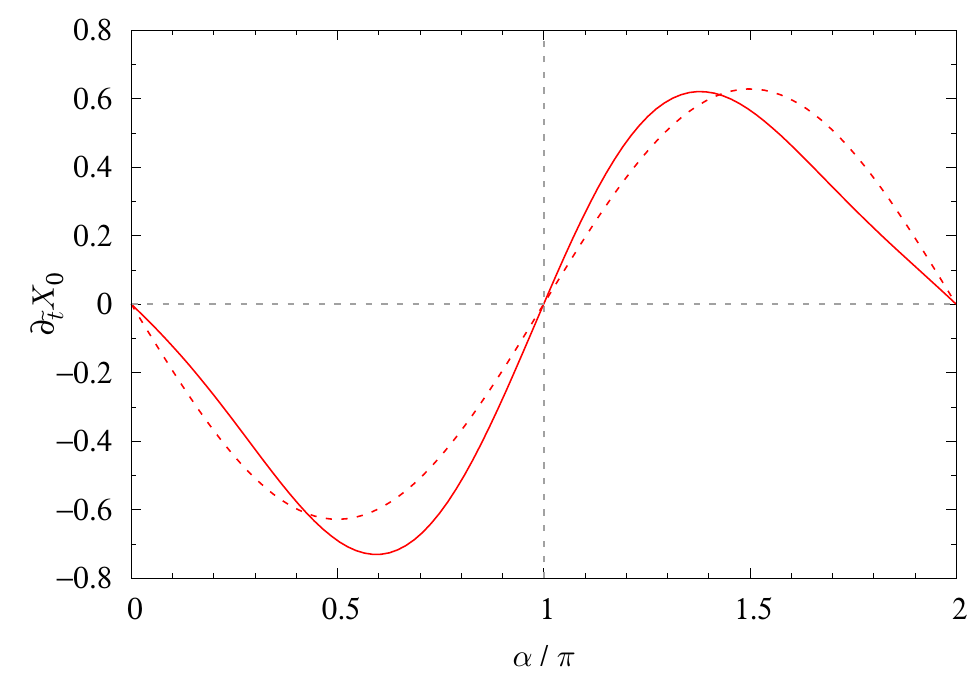}}
    \subfloat[]{\includegraphics[width=0.49\linewidth]{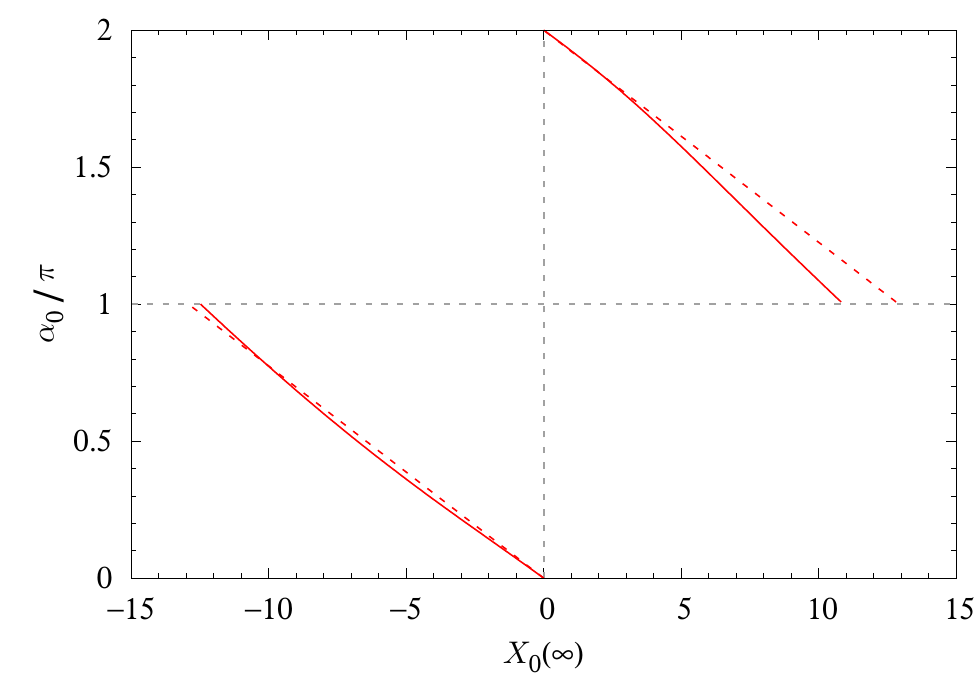}}}
  \mbox{\subfloat[]{\includegraphics[width=0.49\linewidth]{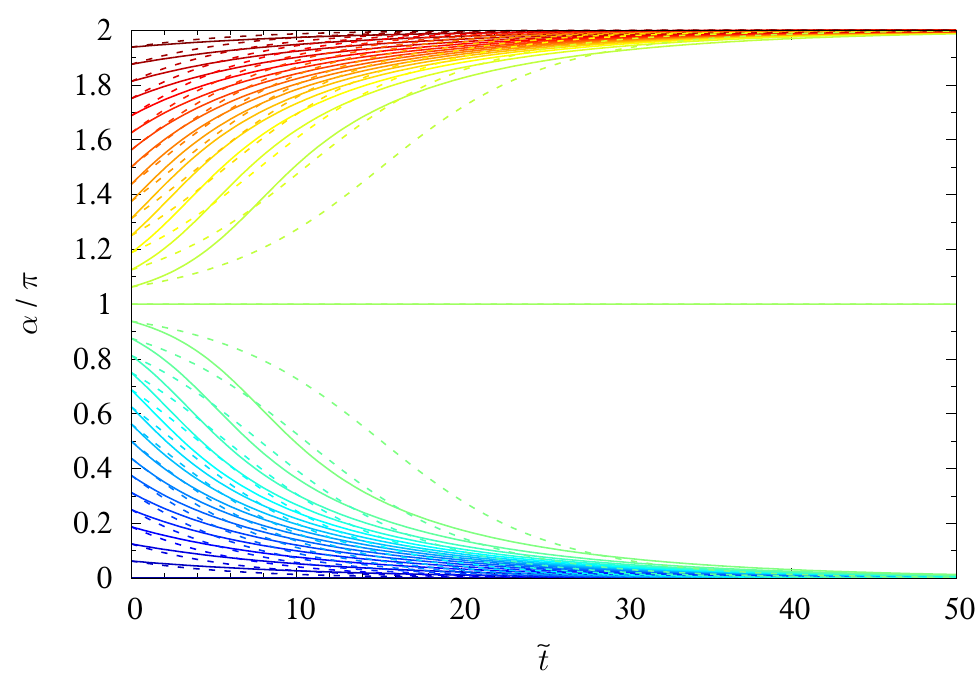}}
    \subfloat[]{\includegraphics[width=0.49\linewidth]{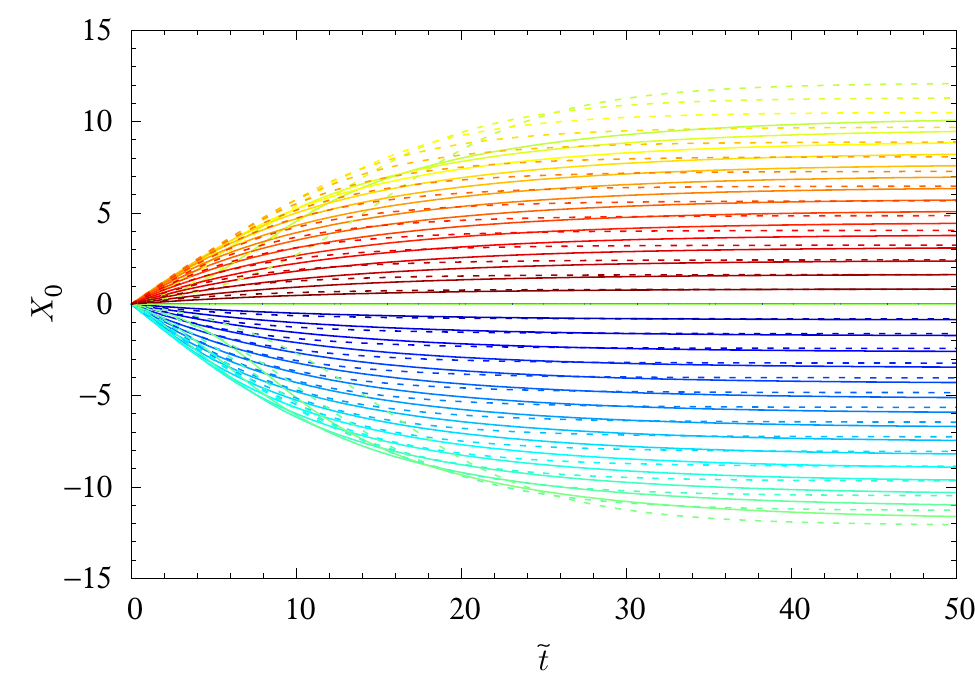}}}
  \caption{Thiele equation dynamics of the DW with a N\'eel DMI.
    (a) The Thiele equation \eqref{eq:Neel_X0t_eff} as a function of
    the DW phase $\alpha$. (b) The final position of the DW,
    $X_0(\infty)$, as a function of the initial value of the DW phase
    $\alpha_0=\alpha(0)$ (the plot is transposed on purpose for easier
    comparison with the phase diagram).
    (c) The DW angle, $\alpha$, and (d) the DW position, $X_0$, as
    functions of time during Thiele (LLG) flow.
    The color coding between panel (c) and (d) is the same: the
    initial phase $\alpha_0$ in panel (d) corresponds to the left-most
    value of $\alpha$ in panel (c).
    The dashed lines correspond to the exact analytic solutions, whereas
    the solid lines are numerical solutions that take into account the 
    demagnetization effect.
    All quantities in this figure are dimensionless and the values of
    the constants are given in Eq.~\eqref{eq:modelparms}.
  }
  \label{fig:movingNeelDWmodulispace}
\end{figure}
Qualitatively, the dynamics in the case of the N\'eel DMI is similar
to the case of the Bloch DMI.
However, we notice that for positive and large final positions
$X_0(\infty)$, the demagnetization effect moves the DW further away in
the Bloch case, but closer to the starting point in the N\'eel case,
see Figs.~\ref{fig:movingDWmodulispace}(b) and
\ref{fig:movingNeelDWmodulispace}(b), respectively. 
The DW phase ($\alpha$) and position ($X_0$) are plotted as functions
of dimensionless time in Figs.~\ref{fig:movingNeelDWmodulispace}(c) and
\ref{fig:movingNeelDWmodulispace}(d), respectively, whereas the
direction of the movement of the DW in the N\'eel DMI case is shown in
Fig.~\ref{fig:movingNeelDWmodulispace}(a).
The exact analytic solutions for $\alpha$ and $X_0$ are shown 
in Fig.~\ref{fig:movingNeelDWmodulispace} with dashed lines and numerical 
solutions for the case incorporating the effect of demagnetization with 
$\eta=0.3$ are shown with solid lines.

\subsection{Phase diagram}\label{sec:phasediagram}

\begin{figure}[!htp]
  \centering
  \includegraphics[width=\linewidth]{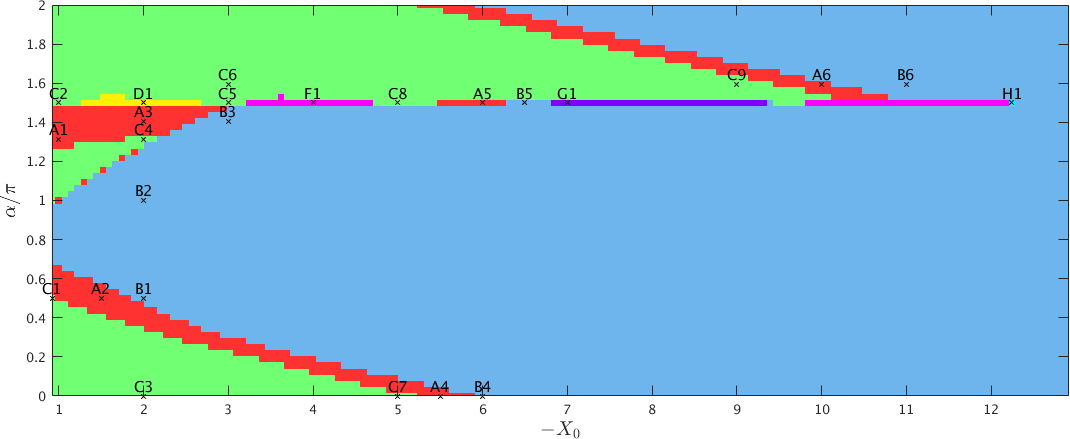}
  \caption{Phase diagram for Bloch DMI \emph{without} demagnetization taken
    into account. The axes correspond to initial values
    for the DW phase ($\alpha$) and position ($X_0$) and the color
    code represents the final state displayed in
    Fig.~\ref{fig:finalstates}.
    Green means creation of a (single) DW-skyrmion, red means
    annihilation of the original skyrmion, and blue means repulsion of
    the original bulk skyrmion.
    The other colors are more exotic final states illustrated in
    Fig.~\ref{fig:finalstates}.
    The labels with corresponding mark (${\scriptscriptstyle\times}$)
    pinpointing the coordinates in the phase diagram, correspond to a
    video of the full simulation that can be found with the label's
    name in the ancillary files.
    The topological charge as a function of time,
    $Q(\tilde{t})$, for the labeled simulations is shown in
    Fig.~\ref{fig:Qt} in Appendix \ref{app:topo_charge}.
    This diagram equally applies to the case of the N\'eel DMI without
    demagnetization by shifting the $\alpha$ axis:
    $\alpha\to\alpha-\pi/2$. 
  }
  \label{fig:phasediagram}
\end{figure}

\begin{figure}[!htp]
  \centering
  \includegraphics[width=\linewidth]{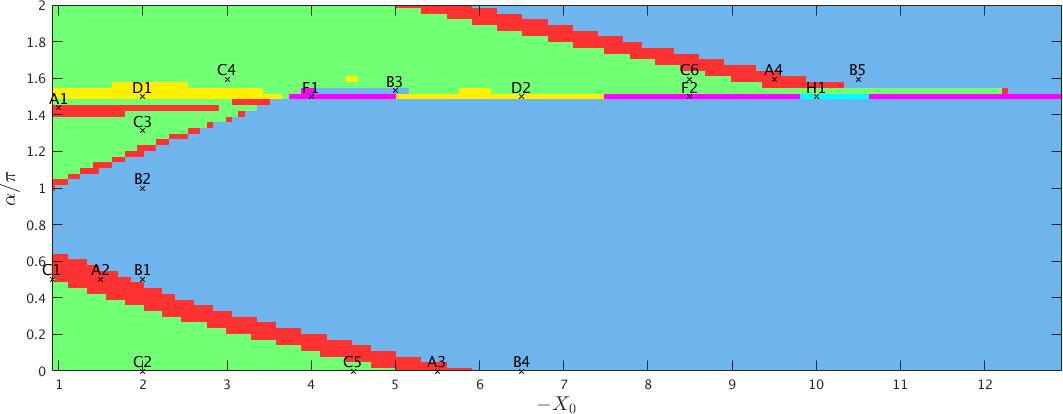}
  \caption{Phase diagram for Bloch DMI \emph{with} demagnetization taken
    into account. The axes correspond to initial values
    for the DW phase ($\alpha$) and position ($X_0$) and the color
    code represents the final state displayed in
    Fig.~\ref{fig:finalstates_Bdm}.
    The labels with corresponding mark (${\scriptscriptstyle\times}$)
    pinpointing the coordinates in the phase diagram, correspond to a
    video of the full simulation that can be found with the label's
    name and the postfix \texttt{\_Bloch\_demag}
    in the ancillary files,
    e.g.~\texttt{C1\_Bloch\_demag.mp4}.
    The topological charge as a function of time,
    $Q(\tilde{t})$, for the labeled simulations is shown in
    Fig.~\ref{fig:Qt_Bdm} in Appendix \ref{app:topo_charge}.
  }
  \label{fig:phasediagram_Bdm}
\end{figure}

Let us now turn to the phase diagram of the experiment with initial
state \eqref{eq:u_composite}, DW phase $\alpha$ and DW position $X_0$.
Evolving full numerical LLG flow of the configuration with said
initial state according to the numerical method of
Sec.~\ref{sec:num_method}, we obtain one of the final states of
Fig.~\ref{fig:finalstates}.
The result for the full phase diagram of final states for a given
pair of initial parameters ($\alpha$, $X_0$) is shown in
Fig.~\ref{fig:phasediagram} for the case of the Bloch DMI without
demagnetization taken into account.
This case equally describes the N\'eel DMI without demagnetization by
shifting $\alpha\to\alpha-\pi/2$.

If we start with the stable DW in the case of the Bloch DMI,
$\alpha=\pi/2$, the three different outcomes depend only on the
distance between the isolated bulk skyrmion and the DW, given by
$|X_0|$:
If the bulk skyrmion is too far away, it is repelled as
explained by the static asymptotic interaction computed in
Ref.~\cite{Gudnason:2024shv}.
If the bulk skyrmion is in a critically
close range to the DW, it suffers a shrinking instability given by the
fact that the bulk skyrmion has a negative DMI energy, whereas the
DW-skyrmion always has a positive DMI energy.
If the bulk skyrmion is caught in the middle, so-to-speak, it has a
small or vanishing DMI energy and is unstable to shrink to a point --
the skyrmion disappears.
If the bulk skyrmion is close enough to the DW, it can successfully be
captured and converted into a DW-skyrmion with positive DMI energy.
These three cases are marked with labels B1, A2 and C1, respectively,
in Fig.~\ref{fig:phasediagram}, which also correspond to simulation
videos available in the ancillary files with the same names.

Moving away from the stable DW phase ($\alpha=\pi/2$), the story is
similar, but the critical distance between the DW and the bulk
skyrmion changes.
In fact, the distance for a successful capture is about twice as large
when using LLG flow as opposed to energy minimization techniques.
This can be understood from the fact that the unstable DW moves
under LLG flow: it moves right (i.e.~toward the bulk skyrmion) for
$\alpha<\pi/2$ and $\alpha>3\pi/2$, whereas it moves left (i.e.~away
from the bulk skyrmion) for $\pi/2<\alpha<3\pi/2$.
This can be understood qualitatively quite well from the Thiele
equation, as shown in Fig.~\ref{fig:movingDWmodulispace}.
This means that apart from the critical distance between the bulk
skyrmion and the DW, the situation is again the same with three
possible outcomes that are determined by the separation distance, see
e.g.~the labels (B4, A4, C7) or (B6, A6, C9) in
Fig.~\ref{fig:phasediagram}.

Two exceptions to this general situation occur: The red area for
$1.2\pi\lesssim\alpha<3\pi/2$ and small $X_0\lesssim3$.
This is actually a successful capture of the bulk skyrmion to become a
DW-skyrmion.
What happens in this case is due to the size of the simulation box,
the choice of Neumann boundary conditions at the endpoints of the DW
and the fact that the LLG flow makes the skyrmion move sufficiently
downwards (in the $-\hat{y}$ direction) that the DW-skyrmion leaves
the simulation box, see the videos A1 and A3 in the ancillary files.

The other and more interesting exception, is the fine-tuned point
$\alpha=3\pi/2$ corresponding to the unstable Bloch DW: this gives
rise to a 1-dimensional Kibble-Zurek mechanism, as first mentioned in
Ref.~\cite{Gudnason:2024shv}.
Although somewhat fine-tuned, this also gives the possibility to
create multiple DW-skyrmions or DW-skyrmion-anti-DW-skyrmion pairs.
We will denote the line in the phase diagram where the 1-dimensional
Kibble-Zurek mechanism takes place as the ``Kibble line''. 
We will discuss this case in the next subsection.

For comparison, we take the effect of demagnetization into account in
Fig.~\ref{fig:phasediagram_Bdm} in the case of the Bloch DMI.
Qualitatively, the phase diagram is quite similar, but the minute
details are slightly different, which can be seen by inspecting
Figs.~\ref{fig:phasediagram} and \ref{fig:phasediagram_Bdm}.
The red area for $1.2\pi\lesssim\alpha<3\pi/2$ and small
$X_0\lesssim3$ has shrunk when demagnetization is taken into account.
This physically probably means that the DW-skyrmion's motion is more
dissipative.

The biggest impact the effect of the demagnetization has in the case
of the Bloch DMI is seen on the Kibble line, which we will discuss in
the next subsection.
It is worth mentioning that the Kibble line does not end with
repulsion (blue color code) at $|X_0|\approx13$, but continues far
beyond the displayed phase diagram with nontrivial creation of
DW-skyrmions.
This is, however, not due to the presence of the bulk magnetic
skyrmion, but an effect of the demagnetization of the intermediate state
of the DW with a Bloch DMI -- seemingly, it is numerically impossible
to fall into the ground state without inducing the Kibble-Zurek mechanism.

\begin{figure}[!htp]
  \centering
  \includegraphics[width=\linewidth]{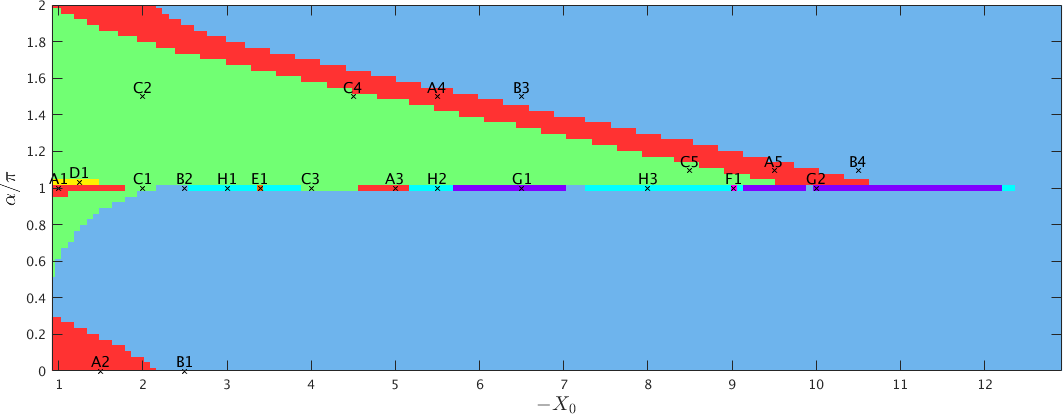}
  \caption{Phase diagram for N\'eel DMI \emph{with} demagnetization taken
    into account. The axes correspond to initial values
    for the DW phase ($\alpha$) and position ($X_0$) and the color
    code represents the final state displayed in
    Fig.~\ref{fig:finalstates_Ndm}.
    The labels with corresponding mark (${\scriptscriptstyle\times}$)
    pinpointing the coordinates in the phase diagram, correspond to a
    video of the full simulation that can be found with the label's
    name and the postfix \texttt{\_Neel\_demag}
    in the ancillary files,
    e.g.~\texttt{C1\_Neel\_demag.mp4}.
    The topological charge as a function of time,
    $Q(\tilde{t})$, for the labeled simulations is shown in
    Fig.~\ref{fig:Qt_Ndm} in Appendix \ref{app:topo_charge}.
  }
  \label{fig:phasediagram_Ndm}
\end{figure}

We finally turn to the case of the N\'eel-type DMI with
demagnetization taken into account, see the phase diagram in
Fig.~\ref{fig:phasediagram_Ndm}.
Clearly, the Kibble line has now changed drastically again and we will
discuss this in the next subsection.
Perhaps the most interesting change from the case without
demagnetization to the case with demagnetization and N\'eel-type DMI,
is that, although for $\pi<\alpha<2\pi$ the situation is as usual a
blue/red/green outcome corresponding to
repulsion/annihilation/DW-skyrmion-creation, for $0<\alpha<\pi$ the
situation is different.
Indeed, for $0<\alpha<0.3\pi$ there is no creation of DW-skyrmions.
This whole red area is characterized by the shrinking instability.
Moreover, for $\pi/2<\alpha<\pi$ the DW-skyrmion creation area (green 
color code) is directly adjacent to the repulsion area (blue color
code): there is seemingly no middle instability between creation and
repulsion, which generically happened in the other cases.

\subsection{The Kibble line}\label{sec:Kibble_line}

\begin{figure}[!htp]
  \centering
  \includegraphics[width=0.8\linewidth]{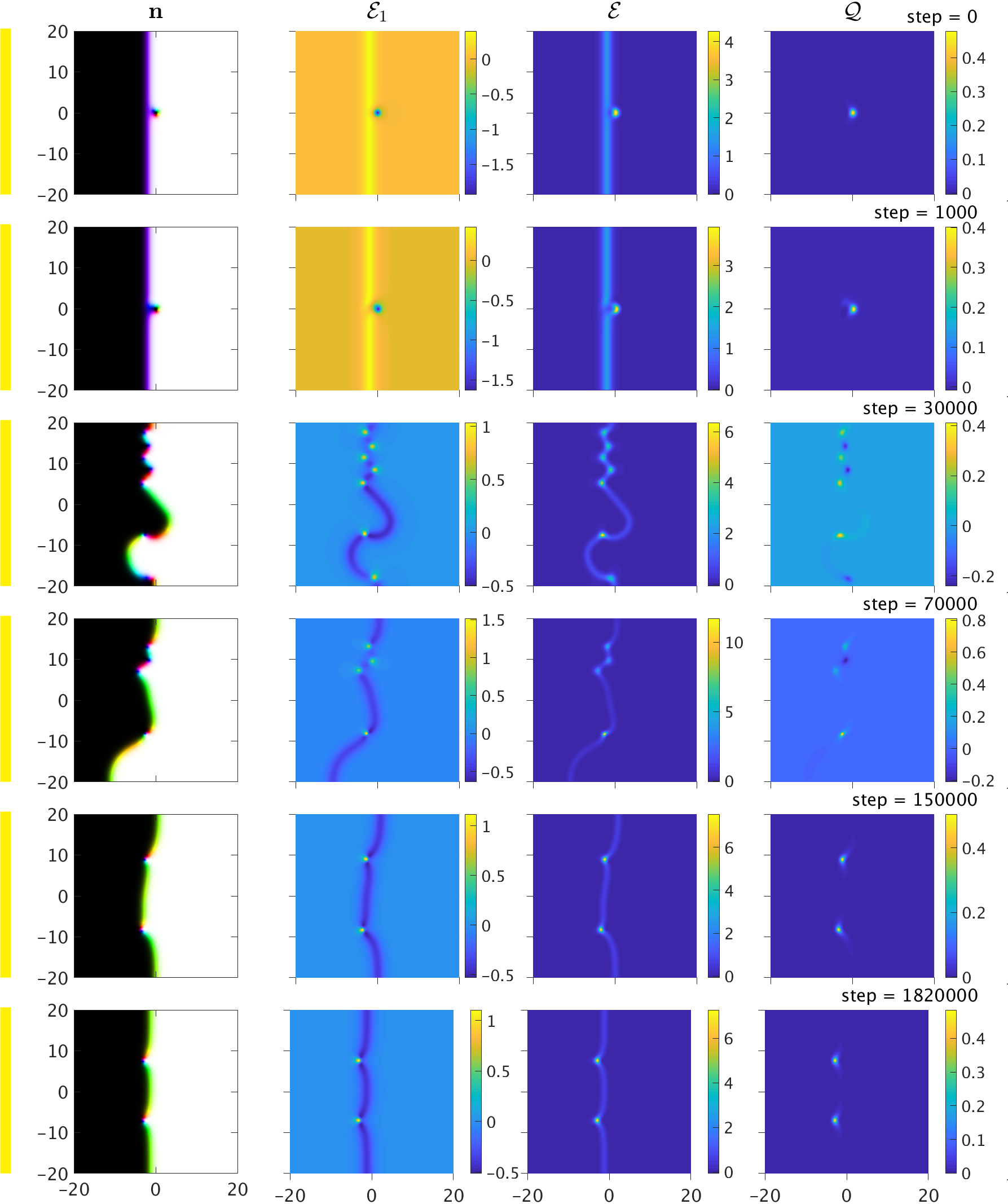}
  \caption{LLG flow of initial configuration D1 in
    Fig.~\ref{fig:phasediagram} in the case of Bloch DMI
    \emph{without} demagnetization.
    The columns of the figure correspond to
    the color code of the final state (i.e.~the last row of
    the figure, see Fig.~\ref{fig:finalstates}),
    the magnetization field,
    the DMI energy density, the total energy density and the
    topological charge density, respectively.
    The rows correspond to the time evolution according to the LLG
    equation with the first and the last row corresponding the initial
    and final state, respectively.
    The intermediate rows correspond to selected snap shots with the
    time step shown above each row.
    Time corresponds to time steps times
    $h_{\tilde{t}}=6\times10^{-4}$ in dimensionless time units.
    The topological charge as a function of time,
    $Q(\tilde{t})$, for this LLG flow is shown in Fig.~\ref{fig:Qt}.
    This particular Kibble line configuration ends up in the yellow
    final state (see Fig.~\ref{fig:finalstates}),
    viz.~a double DW-skyrmion configuration without a
    bulk skyrmion -- the bulk skyrmion has been absorbed into the DW.
  }
  \label{fig:D1}
\end{figure}

\begin{figure}[!htp]
  \centering
  \includegraphics[width=0.8\linewidth]{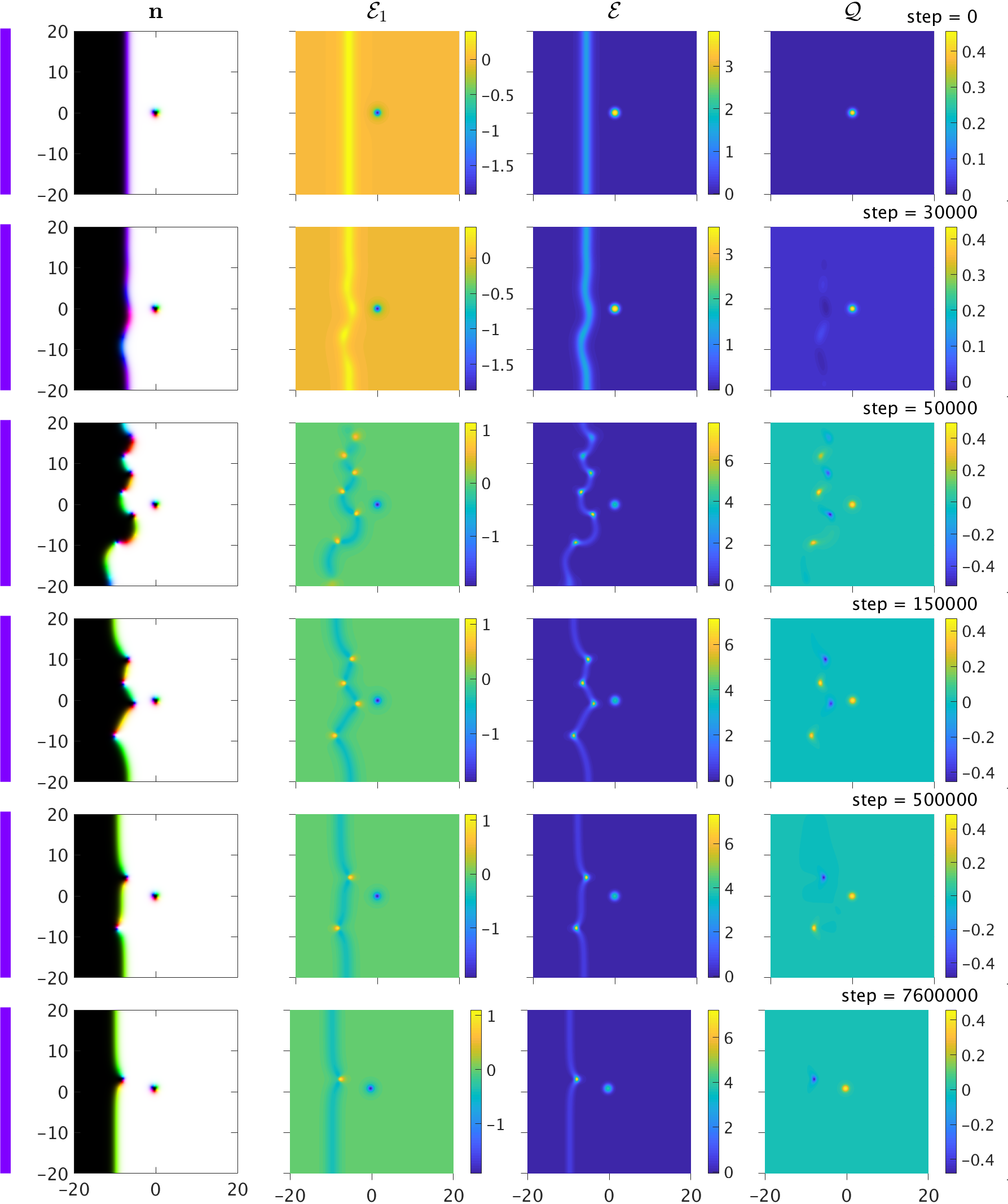}
  \caption{LLG flow of initial configuration G1 in
    Fig.~\ref{fig:phasediagram} in the case of Bloch DMI
    \emph{without} demagnetization.
    The topological charge as a function of time,
    $Q(\tilde{t})$, for this LLG flow is shown in Fig.~\ref{fig:Qt}.
    For details of the figure, see the caption of Fig.~\ref{fig:D1}. 
    This particular Kibble line configuration ends up in the purple
    final state (see Fig.~\ref{fig:finalstates}),
    viz.~an anti-DW-skyrmion configuration with a
    bulk skyrmion -- the bulk skyrmion was never absorbed into the DW
    in this case.
  }
  \label{fig:G1}
\end{figure}

Starting with the Kibble line for the case of the Bloch DMI without
demagnetization (i.e.~$\alpha=3\pi/2$, $|X_0|\lesssim12.2$), see
Fig.~\ref{fig:phasediagram}, we illustrate the LLG dynamics in
Figs.~\ref{fig:D1} and \ref{fig:G1} as representative examples.
The figures are organized in rows at selected time steps as an integer
multiplying the temporal step size $h_{\tilde{t}}=6\times10^{-4}$ (see
Sec.~\ref{sec:num_method}), with the first row showing the initial
condition and the last row the final state.
Because the DW is at the unstable point, a tiny perturbation will make
it fall into the ground state either by $\alpha\to2\pi\to\pi/2$ or by
$\alpha\to\pi\to\pi/2$.
Either way, once a segment starts falling into the ground state, the
symplectic part of the LLG equation causes cusps to appear on the DW
effectively creating a large number of DW-skyrmion-anti-DW-skyrmion
pairs.
Many of them are too close together to survive the LLG flow and will
quickly mutually annihilate, leaving behind only a couple of
DW-skyrmions or a DW-skyrmion-anti-DW-skyrmion pair.

We can classify the outcomes again into three cases: the bulk magnetic 
skyrmion is absorbed into the DW, it is annihilated (shrunk to a
point) or it is repelled.
In every case, the bulk magnetic skyrmion perturbs the DW creating a
number of DW-skyrmions and anti-DW-skyrmions.

If we start from the left (small $|X_0|$) in
Fig.~\ref{fig:phasediagram}, the bulk skyrmion is absorbed into the DW
and two DW-skyrmions are created but one of them is too close to the
boundary of the box and has too much kinetic energy so that it leaves
the simulation area (e.g.~C2 in Fig.~\ref{fig:phasediagram}).
Moving rightwards (larger $|X_0|$) the color code turns yellow meaning
that the two DW-skyrmions remain (e.g.~D1 in
Fig.~\ref{fig:phasediagram}), see Fig.~\ref{fig:D1}.
Moving rightwards again, one of the two DW-skyrmions slips off the
simulation area (e.g.~C5 in Fig.~\ref{fig:phasediagram}).
Continuing right, the bulk magnetic skyrmion is now so far away from
the DW that it does not get absorbed but its perturbation of the DW
creates a DW-skyrmion (e.g.~F1 in Fig.~\ref{fig:phasediagram}).
Moving on in the rightward direction, accidentally an anti-DW-skyrmion
is created in the vicinity of the bulk skyrmion and their fate is
mutual annihilation, leaving behind two DW-skyrmions -- one of which
leaves the simulation area (e.g.~C8 in Fig.~\ref{fig:phasediagram}).
Continuing to the right, the same situation occurs; an
anti-DW-skyrmion annihilates the bulk skyrmion but leaves only a
single DW-skyrmion behind, that nevertheless flows off the simulation
area (e.g.~A5 in Fig.~\ref{fig:phasediagram}).
Moving rightwards, the bulk skyrmion is not absorbed but induces a
DW-skyrmion-anti-DW-skyrmion pair that annihilates itself (e.g.~B5
in Fig.~\ref{fig:phasediagram}).
Continuing to the right, the situation is generally that the bulk
skyrmion is not absorbed by the DW, but it induces a
DW-skyrmion-anti-skyrmion pair. Due to the finite sized simulation
area the DW-skyrmion first slips off (e.g.~G1 in
Fig.~\ref{fig:phasediagram}, see Fig.~\ref{fig:G1}) and later the
anti-DW-skyrmion 
slips off, whereas finally they remain in the DW (e.g.~H1 in
Fig.~\ref{fig:phasediagram}).
After that the bulk skyrmion is so far away that it does not get
absorbed nor does it induce topological solitons on the DW (to the
right of H1 in Fig.~\ref{fig:phasediagram}). 

\begin{figure}[!htp]
  \centering
  \includegraphics[width=\linewidth]{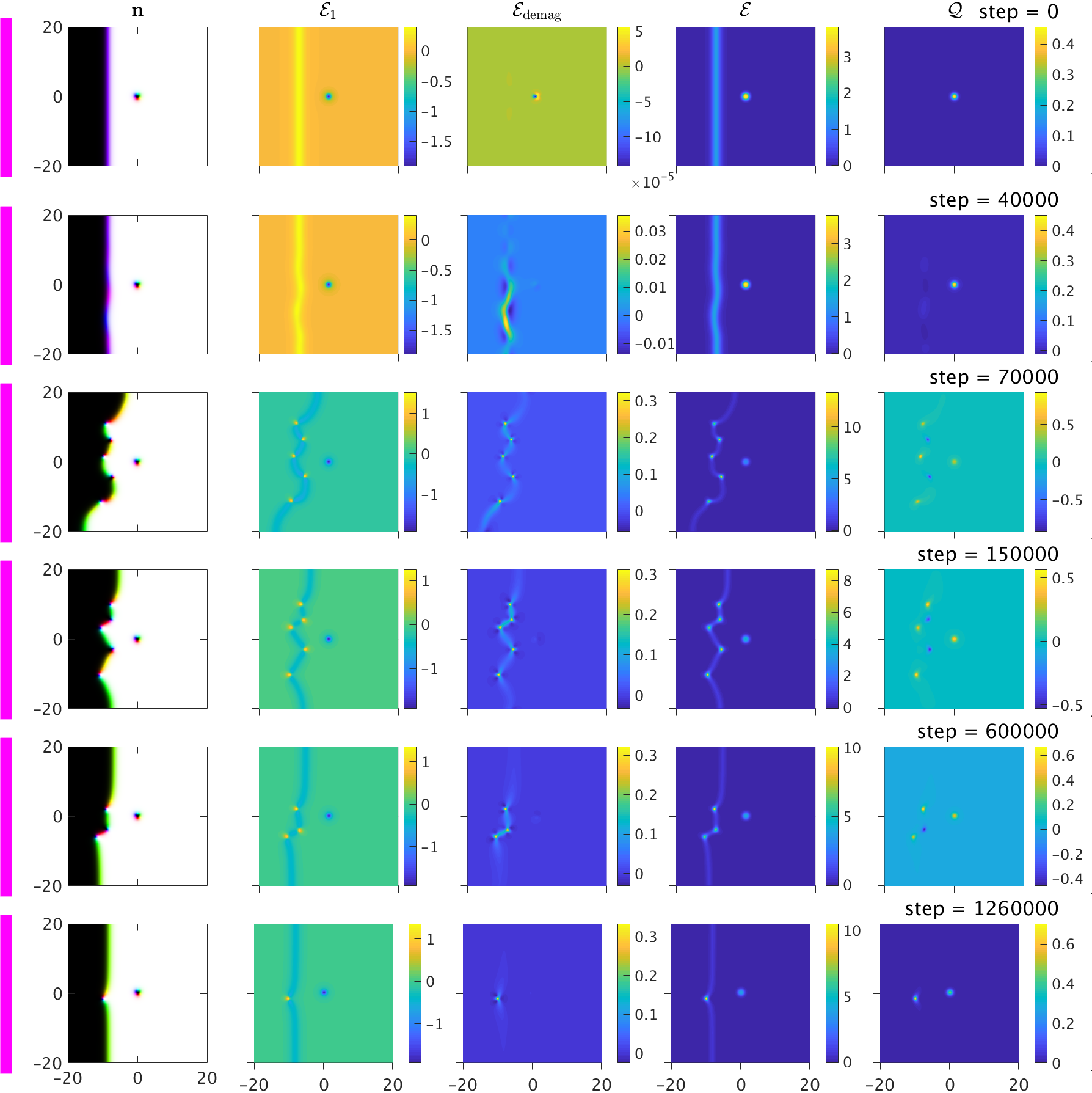}
  \caption{LLG flow of initial configuration F2 in
    Fig.~\ref{fig:phasediagram_Bdm} in the case of Bloch DMI
    \emph{with} the demagnetization effect taken into account.
    The columns of the figure correspond to
    the color code of the final state (i.e.~the last row of
    the figure, see Fig.~\ref{fig:finalstates_Bdm}),
    the magnetization field,
    the DMI energy density, the demagnetization energy density, the
    total energy density and the 
    topological charge density, respectively.
    The topological charge as a function of time,
    $Q(\tilde{t})$, for this LLG flow is shown in Fig.~\ref{fig:Qt_Bdm}.
    This particular Kibble line configuration ends up in the magenta
    final state (see Fig.~\ref{fig:finalstates_Bdm}),
    viz.~a DW-skyrmion configuration with a
    bulk skyrmion -- the bulk skyrmion was never absorbed into the DW
    in this case.
  }
  \label{fig:F2_Bdm}
\end{figure}

\begin{figure}[!htp]
  \centering
  \includegraphics[width=\linewidth]{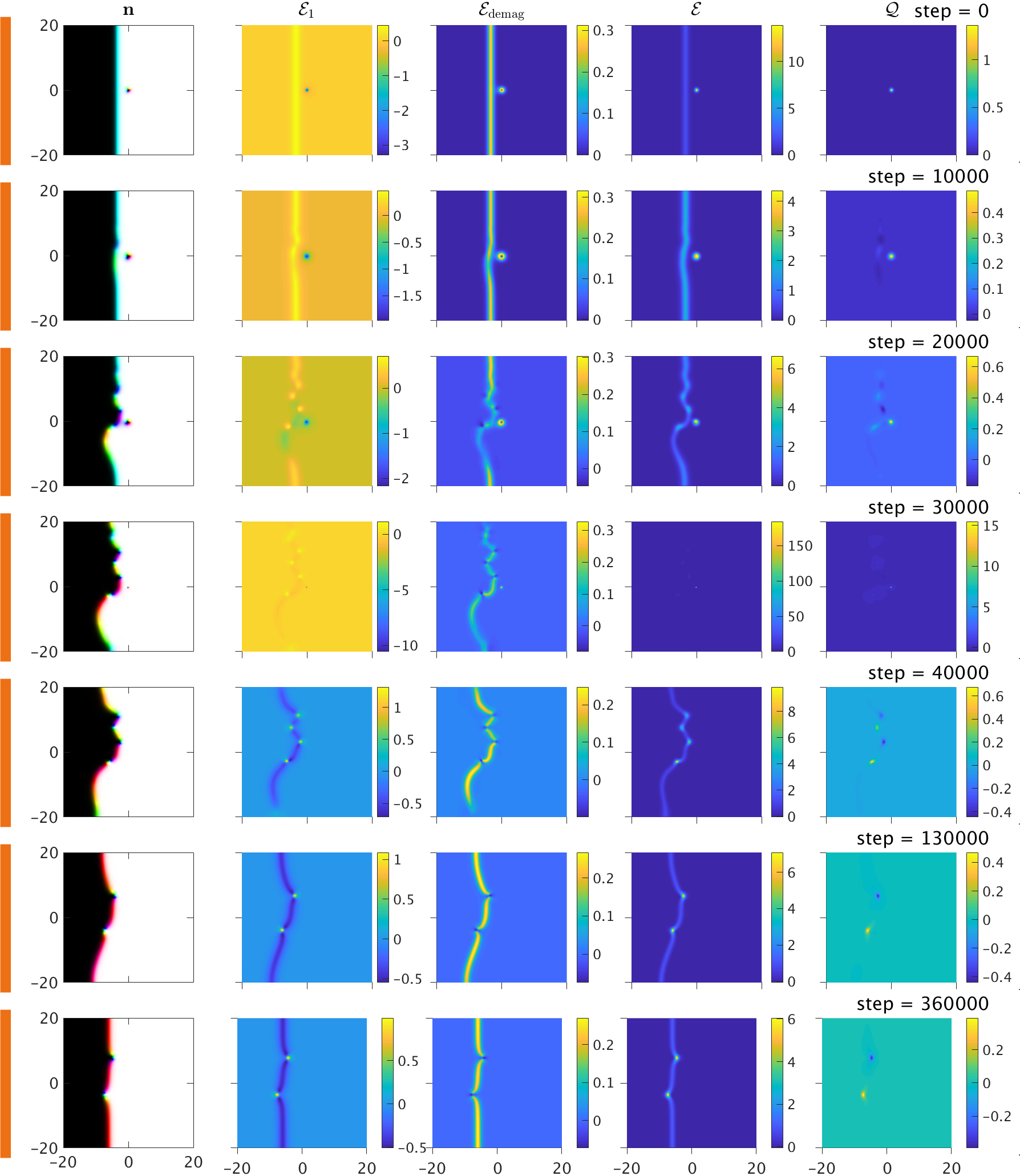}
  \caption{LLG flow of initial configuration E1 in
    Fig.~\ref{fig:phasediagram_Ndm} in the case of N\'eel DMI
    \emph{with} the demagnetization effect taken into account.
    For details of the figure, see the caption of
    Fig.~\ref{fig:F2_Bdm}.
    The topological charge as a function of time,
    $Q(\tilde{t})$, for this LLG flow is shown in Fig.~\ref{fig:Qt_Ndm}.
    This particular Kibble line configuration ends up in the orange
    final state (see Fig.~\ref{fig:finalstates_Ndm}), viz.~a
    DW-skyrmion-anti-DW-skyrmion pair
    \emph{without} a bulk skyrmion -- the bulk skyrmion suffered the
    shrinking instability, as can be seen from rows 3-5.
  }
  \label{fig:E1_Ndm}
\end{figure}

We will now turn to the Kibble line in the case of the Bloch DMI
\emph{with} the demagnetization effect taken into account and as an example
thereof, we illustrate the time evolution during LLG flow of the point
F2 of Fig.~\ref{fig:phasediagram_Bdm} in Fig.~\ref{fig:F2_Bdm}.
We start by noticing that the demagnetization energy vanishes for the
initial configuration (it is of
order $10^{-4}$ due to the fact that the superposition of two solitons
is not an exact solution to the equations of motion), see the top row of Fig.~\ref{fig:F2_Bdm}.
Most of the general story from above follows through.
It is, however, interesting to notice that although both the stable
and the unstable DW in the Bloch DMI case have vanishing
demagnetization energy, the intermediate DW state rotating from an
unstable fixed point to the ground state \emph{does} have a
nonvanishing demagnetization energy.
This fact perturbs the LLG flow.
Qualitatively, the Kibble-Zurek mechanism still happens and in the
illustrated case of Fig.~\ref{fig:F2_Bdm}, the bulk magnetic skyrmion
does not get absorbed into the DW, it induces a number of
DW-skyrmion-anti-DW-skyrmion pairs and at step 600,000 the DW has one
pair and a leftover DW-skyrmion (topmost one), which is the one that
survives on the domain wall.
The DW-skyrmion-anti-DW-skyrmion pair is simply not separated enough
to withstand annihilation.
The end result is a DW-skyrmion and a bulk magnetic skyrmion, a final
state with color code magenta in Fig.~\ref{fig:phasediagram_Bdm}
(marked as F2 in the figure).
We notice that the DW-skyrmion has a nonvanishing demagnetization
energy -- even though the empty DW and the isolated bulk magnetic
skyrmion do not.

Switching the Bloch DMI for the N\'eel DMI \emph{with} the
demagnetization effect taken into account, we illustrate the time
evolution during LLG flow of the point E1 of
Fig.~\ref{fig:phasediagram_Ndm} in Fig.~\ref{fig:E1_Ndm}.
Now the demagnetization effect is fully at play -- every constituent
soliton feels the effect of the demagnetization field, even the empty
DW (see Sec.~\ref{sec:const_soliton_Neel}).
The bulk skyrmion clearly perturbs the DW, inducing the Kibble-Zurek
mechanism on the DW which creates a number of
DW-skyrmion-anti-DW-skyrmions.
Interestingly, in this case, the bulk magnetic skyrmion is neither
absorbed nor repelled from the DW, but it is shrunk to a point, see
rows 3-5 in Fig.~\ref{fig:E1_Ndm}.
The DW-skyrmion-anti-DW-skyrmions on the DW evolve according to the
LLG flow and a single pair is left behind with a sufficiently large
separation distance so that it does not self-annihilate.
The final state is exactly one DW-skyrmion-anti-DW-skyrmion pair,
which has color code orange in Fig.~\ref{fig:phasediagram_Ndm} (marked
as E1 in the figure).

\section{Conclusion and outlook}\label{sec:conclusion}

In this paper, we have continued the study of the possibility of
absorbing an isolated bulk magnetic skyrmion into a DW, which we
started in Ref.~\cite{Gudnason:2024shv}.
In Ref.~\cite{Gudnason:2024shv} we only considered the Bloch DMI
without taking into account the effect of demagnetization and we found
the nearest final states using the far more efficient energy
minimization method called arrested Newton flow.
If there were only one ground state, the LLG equation and the arrested
Newton flow (aNF) would both find the ground state, taking different
amounts 
of time (the aNF method is approximately two orders of magnitude
faster than the LLG equation with $\alpha_G=0.3$ and more so with a
smaller Gilbert damping). 

In this paper, we have studied the configuration of DW in the anisotropy
potential, which is perturbed out of its ground state and evolves
according to the LLG equation with a physically reasonable Gilbert
damping.
Moreover, we have considered in this paper both the cases of the Bloch and
the N\'eel DMI, which become physically different from each other when
turning on the effect of the demagnetization field.
We have developed a Thiele equation for the intermediate state of the DW and
used it to show the movement of the DW during an LLG flow to its ground state.
Depending on the phase of the DW, the LLG flow pushes the DW right or
left and in case of a bulk magnetic skyrmion sitting on either side,
this has consequences for obtaining a nontrivial interaction between
the two constituent solitons.
After mapping out some possible final states (more will exist with
larger materials), we have explored the phase diagrams for capture,
annihilation and repulsion in all four cases of Bloch DMI and N\'eel DMI
with and without demagnetization (without demagnetization they are
equivalent up to a mapping of their respective magnetization
vectors).
Finally, we have explored the Kibble line, which is the unstable fixed point
of the DW that under LLG flow induces the 1-dimensional Kibble-Zurek
mechanism creating a number of DW-skyrmion-anti-DW-skyrmion pairs.
Many of them annihilate, but depending on minute details and the
material sizes, many possibilities for DW-skyrmion creation can be
realized.

A natural extension of our work, would be to study the scattering of
the bulk magnetic skyrmion with the DW by including currents in the
simulations.
Although we have studied the absorption of a bulk skyrmion into the
DW, we probably have not found the most efficient or experimentally
viable method yet.

We have taken into account the demagnetization field by using a scalar
magnetic potential suitable for hard ferromagnets without currents.
The assumption is that the motion of the solitons is sufficiently
adiabatic that the induced currents are sufficiently small for this
approximation to be physically reasonable.
Of course, if we include currents or scattering of the solitons, one
may have to use the full vector potential for the magnetic field as
well as the electric potential.
We will leave such extensions for future studies. 

Recently, 3-dimensional structures in chiral ferromagnetic systems
have been discovered theoretically, in particular a composite soliton
consisting of a magnetic skyrmion string attached to a N\'eel-type DW,
i.e.~a DW with an $S^1$ modulus in its world volume
\cite{Gudnason:2025inp}. 
The DW itself would be unstable to rotate into a Bloch DW -- like the
ones studied in this paper -- but is stabilized by the presence of the
magnetic skyrmion string and the fact that the DW bends with a linear
bending forming a cone-like shape.
This composite soliton junction can appear in periodic arrays forming
a junction lattice \cite{Gudnason:2025inp}.
It would be interesting to study the dynamical formation of such
composite soliton junctions or a soliton-junction lattice phase, according to the LLG evolution from
suitable initial configurations and with the demagnetization effect taken
into account.

\subsection*{Acknowledgments}
S.~B.~G.~thanks the Outstanding Talent Program of Henan University for
partial support. 
This work is supported in part by JSPS KAKENHI [Grants No.~JP23KJ1881
  (Y.~A.), No.~JP22H01221 (M.~N.), 
  JP23K22492 (M.~N.)] and the WPI program
``Sustainability with Knotted Chiral Meta Matter (WPI-SKCM2)'' at
Hiroshima University (M.~N.).

\appendix
\renewcommand{\theequation}{A.\arabic{equation}}
\numberwithin{equation}{section}

\section{Random noise as a trigger for the Kibble-Zurek mechanism}\label{app:random}

\begin{figure}[!htp]
  \centering
  \includegraphics[width=0.8\linewidth]{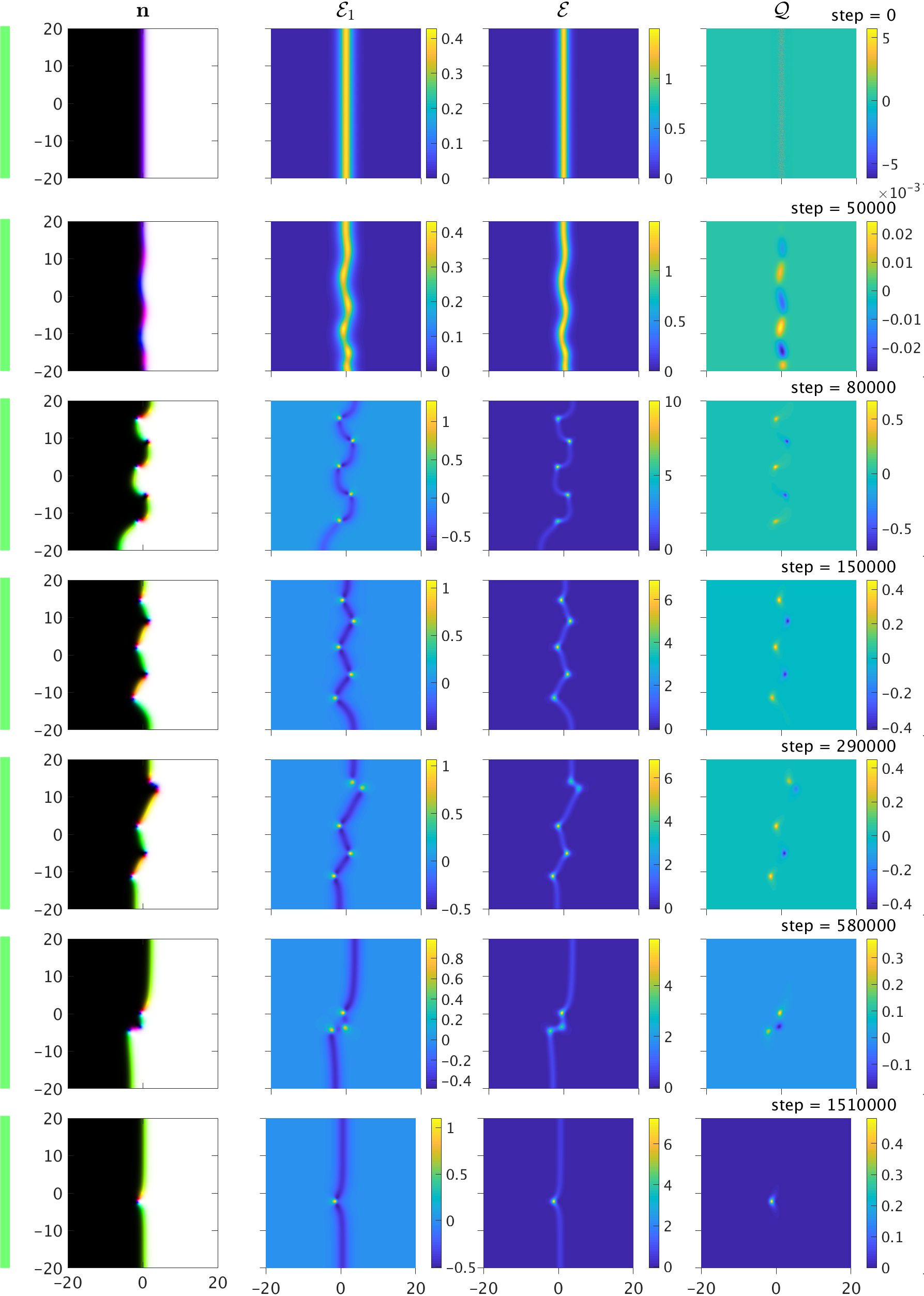}
  \caption{LLG flow of the unstable DW with random noise introduced to
    the DW position variable, $X_0$, in the case of Bloch DMI
    \emph{without} demagnetization. 
    For details of the figure, see the caption of Fig.~\ref{fig:D1}. 
    This particular Kibble line configuration ends up in the green
    final state (see Fig.~\ref{fig:finalstates}),
    viz.~a DW-skyrmion configuration.
  }
  \label{fig:randomnoise}
\end{figure}

In this Appendix, we briefly entertain the possibility of random noise as the trigger for the Kibble-Zurek mechanism to produce a number of DW-skyrmion-anti-DW-skyrmion pairs.
In Fig.~\ref{fig:randomnoise}, we illustrate a simulation where we have generated random noise in the position coordinate $X_0\to X_0+\delta X_0$ with $\delta X_0\in[-0.01,0.01]$ being a random noise distribution along the DW.
From brief numerical investigations, we also find that the random noise applied to the DW position, $X_0$, is more efficient at triggering the Kibble-Zurek mechanism than random noise applied to the DW phase variable, $\alpha$. 

\section{Derrick's theorem}\label{app:Derrick}

Working with the energy functional in dimensionless units, we can write
it as
\beq
\widetilde{E} = e_2 + \kappa e_1 + e_0 + \eta e_{\rm demag},
\eeq
which under the rescaling of the coordinates
$\tilde{\bx}\to\Lambda\tilde{\bx}$ scales as \cite{Leask:2025pdz}
\beq
\widetilde{E} = e_2 + \Lambda\kappa e_1 + \Lambda^2 e_0 + \eta e_{\rm demag},
\eeq
where we have defined the terms
\begin{equation}
e_2 = \frac12\int\tilde{\p}_i\bn\cdot\tilde{\p}_i\bn\;\d^2\tilde{x},\qquad
e_1 = \int\bn\cdot\bd_i\times\tilde{\p}_i\bn\;\d^2\tilde{x},\qquad
e_0 = \frac12\int(1-n_3^2)\;\d^2\tilde{x},\qquad
e_{\rm demag} = \bn\cdot\widetilde{\nabla}\Phi\;\d^2\tilde{x}.
\end{equation}
A loop-hole to Derrick's theorem requires
$\d\widetilde{E}/\d\Lambda=0$ to have a real solution for $\Lambda$: 
\beq
\kappa e_1 + 2\Lambda e_0 = 0,
\eeq
Since $\Lambda>0$ and $e_0$ are both positive, a real positive solution
for $\Lambda$ requires $e_1<0$ to be negative for $\kappa>0$ (or
alternatively $e_1>0$ for $\kappa<0$).
The virial law thus requires equilibria (solutions) to satisfy
\beq
\kappa e_1 = -2e_0.
\eeq

Derrick's theorem states that no finite-energy soliton exists in more
than one spatial dimension, barring that the energy functional
consists of a kinetic term and a potential term \cite{Derrick:1964ww}.
The loop-hole for magnetic skyrmions is that the DMI energy is (can be)
negative and is not (classically) conformally invariant (like the
kinetic energy is in two spatial dimensions).
Indeed, the shrinking instability that we observe is due to the DMI energy being positive (in the situation where the skyrmion is close to, but not inside the DW).

\section{Topological charge monitoring}\label{app:topo_charge}

In this Appendix, we illustrate the topological charge as a
function of dimensionless time, $Q(\tilde{t})$, for the labeled
points in the phase diagrams, which also correspond to the simulation
videos in the ancillary files.

\begin{figure}[!htp]
  \centering
  \mbox{\subfloat[Annihilation]{\includegraphics[width=0.49\linewidth]{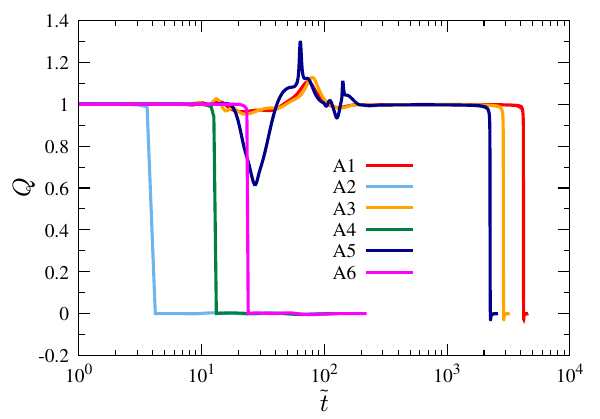}}
    \subfloat[Bounce (repulsion)]{\includegraphics[width=0.49\linewidth]{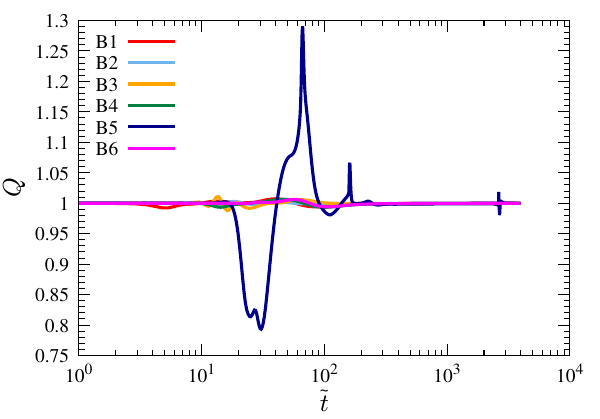}}}
  \mbox{\subfloat[Creation]{\includegraphics[width=0.49\linewidth]{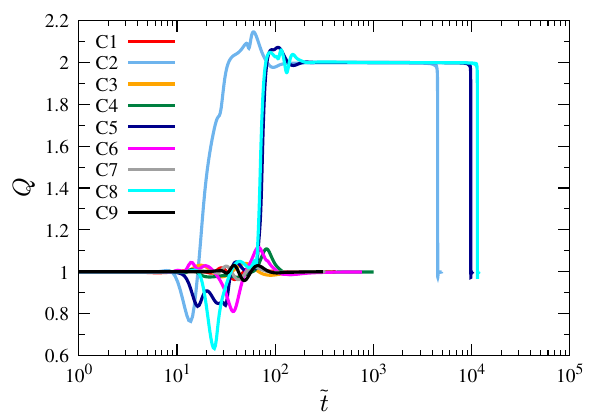}}
    \subfloat[The Kibble line]{\includegraphics[width=0.49\linewidth]{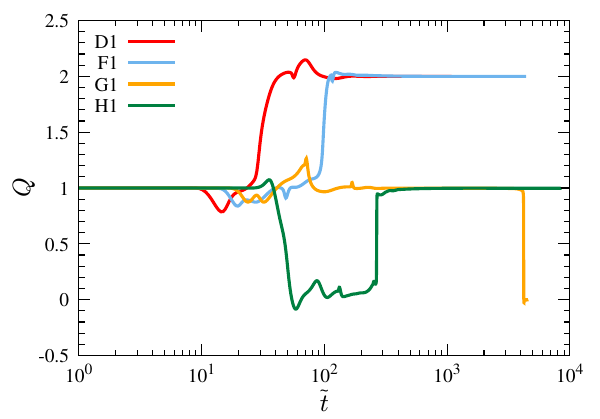}}}
  \caption{
    Topological charge as a function of time in dimensionless
    units, $Q(\tilde{t})$, for the solutions labeled in
    Fig.~\ref{fig:phasediagram} in the case of Bloch DMI
    \emph{without} demagnetization.
    The panels illustrate (a) the annihilations, (b) the repulsions,
    (c) the creations and (d) the Kibble line.
  }
  \label{fig:Qt}
\end{figure}

\begin{figure}[!htp]
  \centering
  \mbox{\subfloat[Annihilation]{\includegraphics[width=0.49\linewidth]{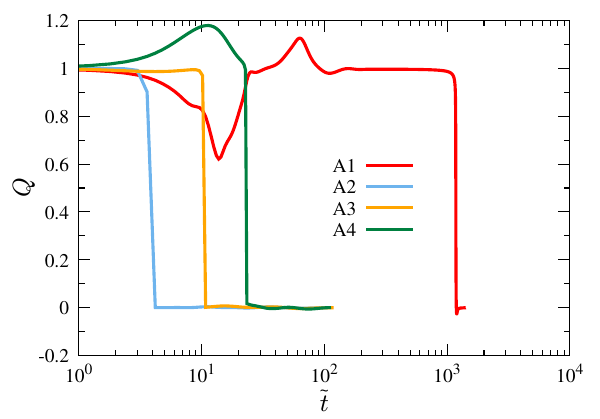}}
    \subfloat[Bounce (repulsion)]{\includegraphics[width=0.49\linewidth]{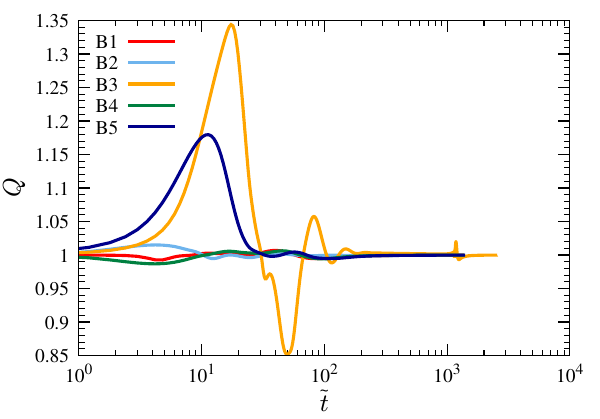}}}
  \mbox{\subfloat[Creation]{\includegraphics[width=0.49\linewidth]{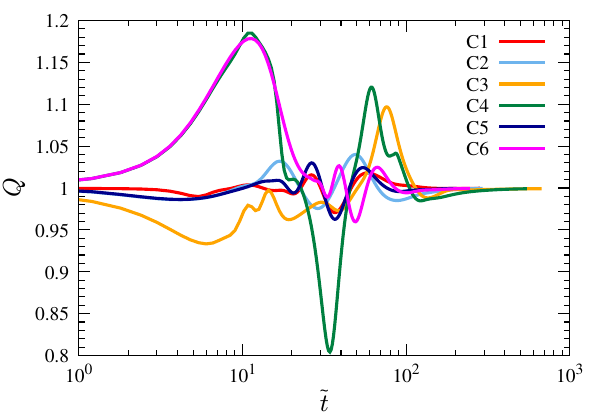}}
    \subfloat[The Kibble line]{\includegraphics[width=0.49\linewidth]{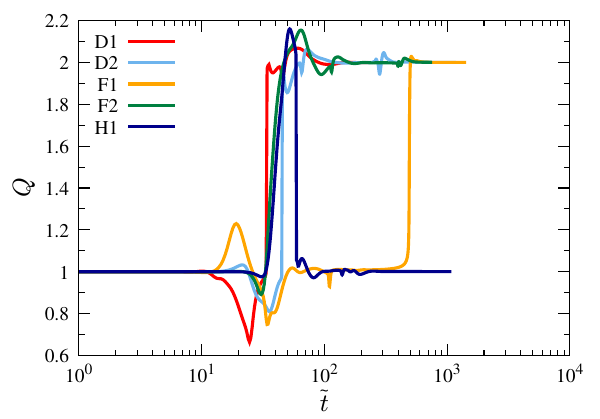}}}
  \caption{
    Topological charge as a function of time in dimensionless
    units, $Q(\tilde{t})$, for the solutions labeled in
    Fig.~\ref{fig:phasediagram_Bdm} in the case of Bloch DMI
    \emph{with} demagnetization.
    The panels illustrate (a) the annihilations, (b) the repulsions,
    (c) the creations and (d) the Kibble line.
  }
  \label{fig:Qt_Bdm}
\end{figure}

\begin{figure}[!htp]
  \centering
  \mbox{\subfloat[Annihilation]{\includegraphics[width=0.49\linewidth]{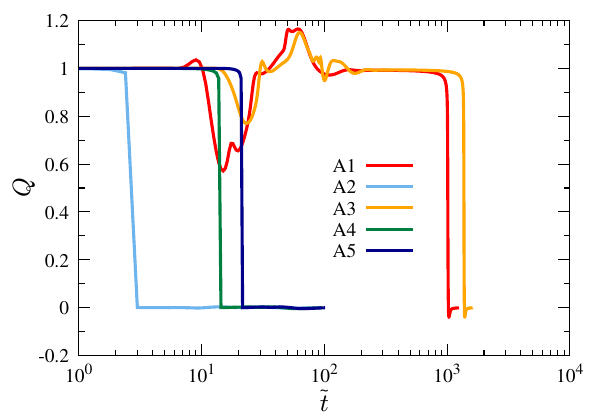}}
    \subfloat[Bounce (repulsion)]{\includegraphics[width=0.49\linewidth]{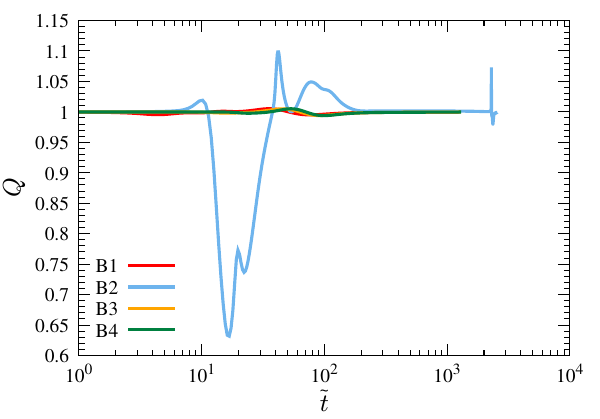}}}
  \mbox{\subfloat[Creation]{\includegraphics[width=0.49\linewidth]{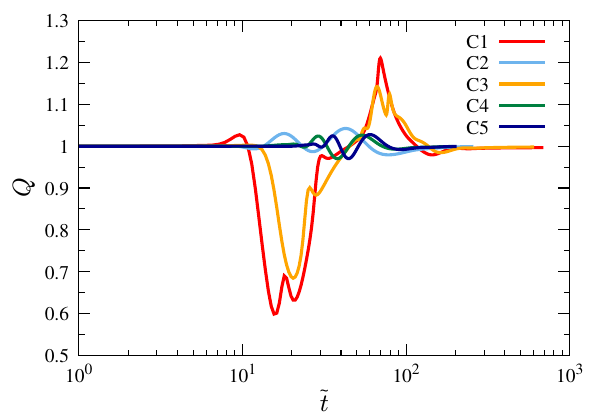}}
    \subfloat[The Kibble line]{\includegraphics[width=0.49\linewidth]{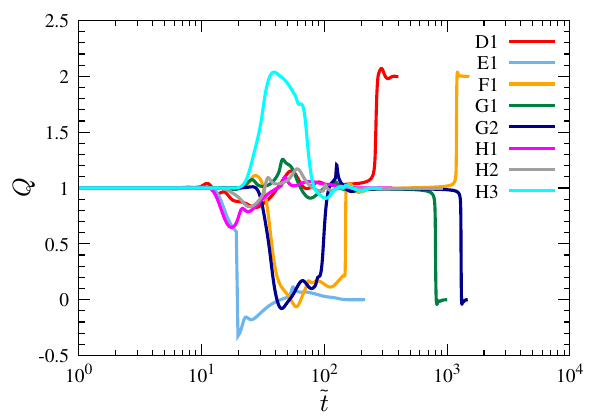}}}
  \caption{
    Topological charge as a function of time in dimensionless
    units, $Q(\tilde{t})$, for the solutions labeled in
    Fig.~\ref{fig:phasediagram_Ndm} in the case of N\'eel DMI
    \emph{with} demagnetization.
    The panels illustrate (a) the annihilations, (b) the repulsions,
    (c) the creations and (d) the Kibble line.
  }
  \label{fig:Qt_Ndm}
\end{figure}

Starting with the Bloch/N\'eel case \emph{without} demagnetization,
we show the integrated topological charge $Q(\tilde{t})$ for all
labeled points in the phase diagram, Fig.~\ref{fig:phasediagram}.
The six annihilations are shown in Fig.~\ref{fig:Qt}a with
A2, A4 and A6 happening quickly, whereas A1, A3 and A5 take two or more
orders of magnitude more time.
Indeed, two different effects are at play here: A2, A4 and A6 are
erased by the shrinking instability described in
Sec.~\ref{sec:phasediagram} and in Ref.~\cite{Gudnason:2024shv}.
On the other hand, A1, A3 and A5 are examples of creation of a
DW-skyrmion which, however, flows out of the simulation area and
hence is marked as ``annihilation''.
Similar situations are seen in the cases \emph{with} demagnetization
in Fig.~\ref{fig:Qt_Bdm}a (Bloch DMI case) and
Fig.~\ref{fig:Qt_Ndm}a (N\'eel DMI case).

Fig.~\ref{fig:Qt}b shows the repulsions or bounces, which is
the natural situation due to the repulsive force between the
Bloch-DW and the magnetic skyrmion in the ground state, see
Ref.~\cite{Gudnason:2024shv}.
One particularly wiggly case is B5, which at time
$\tilde{t}\approx 20$-$100$ shows a great wiggle with $Q$ varying from
$0.8$ to $1.3$: this is the single isolated magnetic skyrmion
inducing the Kibble-Zurek mechanism on the DW, whereas the small
wiggle at $\tilde{t}\approx2500$ is the last
DW-skyrmion-anti-DW-skyrmion pair annihilating due to being at too
close proximity to one another.
In the LLG flows that do now show wiggles in the $Q(\tilde{t})$
graph the skyrmion simply bounces back from the DW.
Similar situations are seen in the cases \emph{with} demagnetization
in Fig.~\ref{fig:Qt_Bdm}b (Bloch DMI case) and
Fig.~\ref{fig:Qt_Ndm}b (N\'eel DMI case).

Fig.~\ref{fig:Qt}c shows the creation of DW-skyrmions, which
in the standard case happens by the skyrmion being close enough to
the DW not to experience the shrinking instability (C1, C3, and C4) 
or by the DW moving into the skyrmion (C7 and C9) -- either way this
results in the creation of a DW-skyrmion if the Bloch DW is not in
its ground state (which leads instead to repulsion of the skyrmion).
In the case of C6, the skyrmion is absorbed into the DW forming a
DW-skyrmion but also inducing the Kibble-Zurek mechanism, which
creates a DW-skyrmion-anti-DW-skyrmion pair which, however,
annihilates itself leaving the original DW-skyrmion behind.
Finally, there are three exotic cases: C2, C5 and C8, where the
Kibble-Zurek mechanism is induced with two DW-skyrmions being
created ($Q=2$); this is the yellow final state in
Fig.~\ref{fig:finalstates}.
The reason why they are marked as the creation of single
DW-skyrmions is that one of the two DW-skyrmions flows out of the
simulation area -- indeed, a larger simulation area would have a
$Q=2$ DW-skyrmion final state for these initial conditions.
More exotically, in the case of C8 the magnetic skyrmion is never
absorbed into the DW, but it induces the Kibble-Zurek mechanism that
creates an anti-DW-skyrmion on the DW that annihilates the original
bulk skyrmion.
Similar situations are seen in the cases \emph{with} demagnetization
in Fig.~\ref{fig:Qt_Bdm}c (Bloch DMI case) and
Fig.~\ref{fig:Qt_Ndm}c (N\'eel DMI case), except for the creation of
$Q=2$ DW-skyrmions, which does not happen in the cases shown with
demagnetization.

Finally, we arrive at the Kibble line cases with the
topological charge $Q(\tilde{t})$ shown in Fig.~\ref{fig:Qt}d.
In all cases the Kibble-Zurek mechanism is induced by the magnetic
skyrmion being in close enough proximity to the unstable DW.
The final states are shown in Fig.~\ref{fig:finalstates}.
Similar situations are seen in the cases \emph{with} demagnetization
in Fig.~\ref{fig:Qt_Bdm}d (Bloch DMI case) and
Fig.~\ref{fig:Qt_Ndm}d (N\'eel DMI case).
In all cases, the dynamics is chaotic and for more details and
discussion, see Sec.~\ref{sec:Kibble_line}. 

\bibliographystyle{apsrev4-1}
\bibliography{references}

\end{document}